\documentclass[fleqn,usenatbib]{mnras}

\usepackage[T1]{fontenc}
\usepackage{ae,aecompl}

\usepackage{graphicx}	
\usepackage{amsmath}	
\usepackage{amssymb}	

\usepackage{longtable,rotating}			

\title[The evolution of UDGs in galaxy clusters]{The evolution of ultra-diffuse galaxies in nearby galaxy clusters from the Kapteyn IAC WEAVE INT Clusters Survey (\texttt{KIWICS})}

\author[Pavel E. Mancera Pi\~na et al.]{Pavel E. Mancera Pi\~na$^{1,2}$\thanks{e-mail: pavel@astro.rug.nl},
J. A. L. Aguerri$^{3}$, Reynier F. Peletier$^{1}$, Aku Venhola${^{1,4}}$,\newauthor Scott Trager$^{1}$ and Nelvy Choque Challapa$^{1}$ \\
\\
$^{1}$Kapteyn Astronomical Institute, University of Groningen, Landleven 12, 9747 AD, Groningen, The Netherlands\\
$^{2}$ASTRON, the Netherlands Institute for Radio Astronomy, Postbus 2, 7990 AA, Dwingeloo, The Netherlands\\
$^{3}$Instituto de Astrof\'isica de Canarias, Calle V\'ia L\'actea S/N, La Laguna, Tenerife, Spain\\
$^{4}$Astronomy Research Unit, University of Oulu, Oulu, FI-90014, Finland
}

\date{Accepted XXX. Received YYY; in original form ZZZ}

\pubyear{2018}

\begin{document}
\label{firstpage}
\pagerange{\pageref{firstpage}--\pageref{lastpage}}
\maketitle

\begin{abstract}
We study the population of ultra-diffuse galaxies (UDGs) in a set of eight nearby ($z <$ 0.035) galaxy clusters, from the Kapteyn IAC WEAVE INT Clusters Survey (\texttt{KIWICS}). We report the discovery of 442 UDG candidates in our eight field of views, with 247 of these galaxies lying at projected distances < 1 R$_{200}$ from their host cluster. With the aim of testing theories about their formation, we study the scaling relations of UDGs comparing with different types of galaxies, finding that in the full parameter space they behave as dwarf galaxies and their colors do not seem to correlate with their effective radii. To investigate the influence of the environment on the evolution of UDGs we analyze their structural properties as functions of the projected clustercentric distance and the mass of their host cluster. We find no systematic trends for the stellar mass nor effective radius as function of the projected distance. However, the fraction of blue UDGs seems to be lower towards the center of clusters, and UDGs in the inner and outer regions of clusters have different S\'ersic index and axis ratio distributions. Specifically, the axis ratio distributions of the outer and inner UDGs resemble the axis ratio distributions of, respectively, late-type dwarfs and dwarf ellipticals in the Fornax Cluster suggesting an environmentally-driven evolution and another link between UDGs and dwarf galaxies.  In general our results suggest strong similarities between UDGs and smaller dwarf galaxies in their structural parameters and their transformation within clusters.
\end{abstract}




\section{Introduction}
Ultra-diffuse galaxies (UDGs, \citealt{vandokkum}) are galaxies with effective radius similar to the Milky Way ($R_e \gtrsim$ 1.5 kpc) and extremely low surface brightness ($\mu(g,0) \gtrsim$ 24 mag arcsec$^{-2}$). Their extreme conditions make them important probes for setting constraints on galaxy formation and evolution models, for instance by studying the influence of the cluster environment on the evolution of galaxies.
While low-surface-brightness galaxies (LSB) have been studied for a long time (e.g. \citealt{sandage}; \citealt{impey}; \citealt{bothun}; \citealt{dalcanton}, but see also \citealt{conselice}), recently, with sensitive detectors in large telescopes imaging large areas, it has been possible to perform more and deeper systematic studies (but see also \citealt{reviewlsb}). After the work by \cite{vandokkum}, several works have shown the ubiquity of UDGs in low- and high-density environments, from galaxy clusters (e.g., \citealt{vandokkum}; \citealt{vanderburg}; \citealt{roman}; \citealt{Aku}), to groups (e.g. \citealt{roman2}; \citealt{cohen}), and even in the field (e.g. \citealt{dalcanton}; \citealt{bellazzini}; \citealt{greco}). Furthermore, an analogous HI-rich population has been detected (e.g. \citealt{leisman}; \citealt{jones}.)

Apart from their low luminosities and large sizes, UDGs show a range in color, although in clusters most of them have red colors indicative of old stellar populations (e.g. \citealt{ferre}; \citealt{ruizlara}), light profiles close to exponential, relatively high axis ratios and modest ($\sim$10$^8$M$_\odot$) stellar masses (e.g. \citealt{vandokkum}; \citealt{roman}; \citealt{roman2}; \citealt{vanderburg}; \citealt{Aku})

Their characteristics pose the question whether they are failed L$^\star$ galaxies that did not build up their expected stellar mass or if they are dwarf galaxies with normal stellar mass but unusually large size. The ultimate test to answer this question is to accurately measure their total masses, but given their faintness this is extremely challenging. While a few UDGs have been measured to have massive, Milky Way-sized dark matter haloes (e.g. \citealt{vandokkum2,beasley,toloba}), most of them are observed to inhabit dwarf-like ones (e.g. \citealt{beasleyytrujillo}; \citealt{ferre}) and the subhalo mass function of UDGs \citep{amoriscohaloes} also seems to favor this scenario.

Most of the works trying to explain the origins of UDGs (and classical LSBs) use a framework in which they are dwarfs (e.g. \citealt{bothun}; \citealt{reviewlsb} and references therein, \citealt{amorisco}; \citealt{dicintio}). For instance, focusing on the UDGs, \cite{amorisco} demonstrated that they can be the outcome of dwarf galaxies inhabiting high-spin haloes (but see also \citealt{dalcantondisks}). This high-spin of the halo would then translate into a high-angular momentum (see also \cite{posti}, for a case in which UDGs are also high-angular momentum dwarfs, but non necessarily inhabiting a high-spin halo), making the dwarfs become larger and thus lower surface brightness. On the other hand, \cite{dicintio}, using the \texttt{NIHAO} simulations \citep{nihao}, demonstrated that UDGs can be the outcome of isolated dwarf galaxies that become larger due to strong gas feedback-driven outflows, generated by extended star formation histories (SFHs), regardless of the rotational velocity of their haloes. \cite{baushev} suggested that a fraction of the UDGs could be produced by galaxy collisions in the centres of clusters, while recently, \cite{Aku} have also suggested that tidal interactions in clusters could be the cause of the largest UDGs (see also \citealt{carleton}), similar to \cite{mihos}, who suggested that UDGs may originate from tidally disrupted dwarfs. Adding to this, \cite{bennet} reported the discovery of a couple of disrupting satellite galaxies with similar characteristic as UDGs, with associated tidal features, although they are not particularly large. The emerging panorama seems to be that different formation mechanisms for UDGs are needed to explain all the observed diversity in their properties (e.g. \citealt{papastergis,leisman,ferre,lim}).

Observational and theoretical studies have also addressed the evolution of UDGs in clusters. \cite{vanderburg}, hereafter vdB+16, studied the abundance and spatial distribution of UDGs in a set of nearby galaxy clusters, discovering a tight relation between the number of UDGs and the mass of their host clusters. In \cite{paperI} we found evidence of this slope being sublinear, favoring a scenario where UDGs preferably form (or survive more easily) in low-mass groups (cf. \citealt{roman2}, hereafter RT17b). RT17b combined photometric colors with spectra of a few UDGs, finding trends of UDGs becoming less massive, smaller and lower S\'ersic index at smaller clustercentric distances, in agreement with a picture of UDGs being dwarfs disrupted due to environmental interactions (a caveat to be taken into account is their relatively small sample, and that in general less massive galaxies have smaller S\'ersic index and sizes, so the mass of the galaxies could be playing a role apart from the environmental effects). Their findings are consistent with an evolutionary scenario of UDGs being dwarfs formed outside clusters with relatively blue colors, that later become redder while being accreted onto the clusters, due to the fading after their star formation is quenched (see also \citealt{yozin} and \citealt{alabi}).
\cite{Aku} also showed that the detailed properties of UDGs in Fornax do not differ significantly from those in Virgo and Coma, which is somewhat surprising, given their very different cluster environments. Regarding UDGs in different environments, in \cite{paperI}, hereafter Paper I, we showed that clusters with different masses have their innermost UDGs at different projected distances, arguing in favor of a scenario where UDGs in the centres of high-mass clusters are more efficiently destroyed.\\

\noindent
With all this in mind, and with the aim of updating the census of UDGs in clusters, study their scaling relations and further investigate the influence of the environment on their properties, we present here our second paper in a series studying the UDG population in a set of nearby galaxy clusters. 
\noindent
The rest of this paper is organized as follows. Section \ref{sec:obs} describes the details of our observational campaign, the data reduction process and our sample selection. Section \ref{sec:detection} delves into the detection of potential UDG candidates and their characterization. The structural properties of our sample and the scaling relations are presented in Section \ref{sec:params}. In Section \ref{sec:evolution} we investigate the effects that cluster environment may have on UDGs, by analyzing the abundance and spatial distribution of the UDGs, as well as the dependence of their structural parameters on the environment. We summarise our main results and conclusions in Section \ref{sec:conclusions}.

Throughout this work we use magnitudes in the AB system and we adopt a $\Lambda$CDM cosmology with $\Omega_\textnormal{m}$ = 0.3, $\Omega_{\Lambda}$ = 0.7 and H$_\textnormal{0}$ = 70 km s$^{-1}$ Mpc$^{-1}$.

\section{Observations, data reduction and sample selection}
\label{sec:obs}
\subsection{Observations}
As preparation for one of the galaxy surveys to be carried out with the WEAVE spectrograph \citep{dalton} to be installed in the William Herschel Telescope at the Observatorio del Roque de los Muchachos, in La Palma, Spain our team is doing a deep photometric survey of galaxy clusters (PIs Peletier \& Aguerri): the Kapteyn IAC WEAVE INT Clusters Survey (\texttt{KIWICS}, Choque Challapa et al., in prep.). This survey, imaging 47 X-ray selected clusters (from \citealt{piffaretti}) in the Northern hemisphere, between 0.02 $\leq$ $z$ $\leq$ 0.04, will be excellent for studying the effects of the environment on galaxy evolution, particularly for dwarfs and LSBs.
The observations for this survey are done using the Wide Field Camera\footnote{\url{http://www.ing.iac.es/astronomy/instruments/wfc/}} (WFC) at the 2.5-m Isaac Newton Telescope (INT) in La Palma. This work is based on a subset of those observations.
We use the $g-$ and $r-$ filters with total integration times, per cluster, of $\sim$1800s and $\sim$5400s, respectively (with single exposures of 210s), imaging several fields for each cluster covering at least 1 R$_{200}$ in projection, but our field of view (FOV) is usually larger (see Figure \ref{fig:positions}). The observational strategy combines short exposure times with large dithers between consecutive frames, which allow us to i) reduce the overheads by deriving background models directly from median combining and stacking the science images; ii) reach deep surface brightness limits keeping a high saturation limit; and iii) make sure that the region has a uniform depth thanks to the overlapping fields between dithers. Our observations were obtained in several runs between 2015 and 2018, but they were always done in the same way. 

\subsection{Data reduction}
As in \cite{Aku}, we use the \texttt{Astro-WISE} \citep{astrowise} environment to reduce our data. The data reduction processes have been explained in great detail in \cite{Aku} and \cite{Aku2}, and so we present here only the main steps. The pipeline described in the references above has been very slightly modified to take into account the differences between ESO's VLT Survey Telescope and the INT, but the general procedure is the same, and will be also explained in Choque Challapa et al., in prep.

Our individual cluster images are first bias-subtracted and flat-fielded, using bias and flat-field images obtained for each night of our observations. Bias and flat frames were collected at the beginning and end of each night; in the case of the master flat frame, it is built by combining dome- and twilight-flats. The pipeline also allows making illumination and de-fringing corrections, but we found no need for this.

After this we apply a background subtraction to our science images. This is essential to separate the light from real sources of that from the background. Most contamination from diffuse scattered light in the telescope and from the atmosphere vanishes in this process because of the dithering done when observing: consecutive integrations will not have the same contamination in the same pixels, so by stacking all the images and taking the average, hot and cold pixels, cosmic rays, and fixed pattern noise will be removed. In practice we make the background for each CCD and for each exposure as follows. We first take a set of science frames and the objects on them are masked using \texttt{SExtractor} \citep{sextractor}; the masked pixels are replaced by a low order polynomial fit. 
In practice, as in \cite{Aku}, we used a grid of 50 $\times$ 50 pixels to estimate the background and masking objects above a threshold of 5$\sigma$. Then, for each frame median values are measured in 96 90$\times$90 pixel boxes and scaled with each other. The scaling factor, $s_f$, is given by:
\begin{equation}
s_f = median \left( \dfrac{m_{1,i}}{m_{2,i}} \right) ~,
\end{equation}
with $m_{1,i}$ each of the 96 medians from the reference \textit{image 1}, and $m_{2,i}$ the medians in \textit{image 2}, which is being scaled. If a frame has more than one third of its area masked it is excluded, as well as frames with large scatter in $m_{1,i}/m_{2,i}$. A datacube is built by staking all the remaining frames, and the background model is calculated pixel-by-pixel by taking the median along the z-axis of the datacube. Finally, this model is subtracted from the final image.

Astrometric calibrations are applied by matching sources in the images with the 2 Micron All-Sky Survey Point Source Catalog (2MASS-PSC, \citealt{2mass}) and fitting the residuals by a second-order polynomial surface. In this step the reduced science frames are re-sampled to a scale of 0.2 arcsec pixel$^{-1}$. The astrometry of our final mosaic has an rms of $\sim$0.2 arcsec when compared with the astrometry of \texttt{SDSS}.
Photometric corrections are derived by comparing the instrumental magnitudes of a set of standard stars observed during each night of the observations with the \texttt{SDSS DR14} catalogue \citep{sdss}. Given the similarity of the INT filters with \texttt{SDSS} filters no color term is needed for the calibration. The mean uncertainties of our photometry are of the order of 0.04 and 0.05 mag in the $r-$ and $g-$band, respectively. In this step atmospheric corrections are also applied.

Finally, all the cluster frames are median-stacked to produce a deep coadded mosaic. A weight map of the mosaic is also created, containing the information about saturated or bad pixels (from the hot- and cold-pixel maps), the noise level and cosmic rays.
The mean depth of the $r-$band images is 29.3 mag arcsec$^{-2}$ when measured at a 3$\sigma$ level in boxes of 10''$\times$10''; this is comparable to the typical depths of the latest literature on UDGs (see \citealt{roman2}).

\subsection{The sample}
While our observational survey is still ongoing in 2018, several clusters are ready for analysis. To select our cluster sample, we give priority to clusters with the lowest possible redshift (to have better resolution), that were observed with the lowest seeing of our sample, cf. Table \ref{tab:sample}, and with the best possible image quality. Additionally, we prefer clusters without strong background and/or foreground substructures, to avoid interlopers as much as possible\footnote{Most of our clusters do not have background substructure until $z >$ 0.15 according to the Sloan Digital Sky Survey (SDSS) and NASA/IPAC Extragalactic Database (NED). While this does not ensure that the background does not contain substructures, it implies that most likely, if present, the substructures should not be strong. Additionally, a galaxy with an angular size that at $z$ = 0.025 implies $R_e$ = 1.5 kpc, would need to have a true size of $R_e \sim$ 8 kpc if at $z$ = 0.15; additionally the surface brightness dimming at that redshift is $\sim$ 0.6 mag arcsec$^{-2}$. Therefore we do not expect a strong presence of interlopers from high-redshift substructures.}. At the same time our intention is to cover a range in mass to have a representative sample. In the end we select eight clusters, making this work one of the largest sample of clusters in which UDGs have been studied hitherto, along with vdB+16.

\begin{figure}
\centering
\includegraphics[scale=0.5]{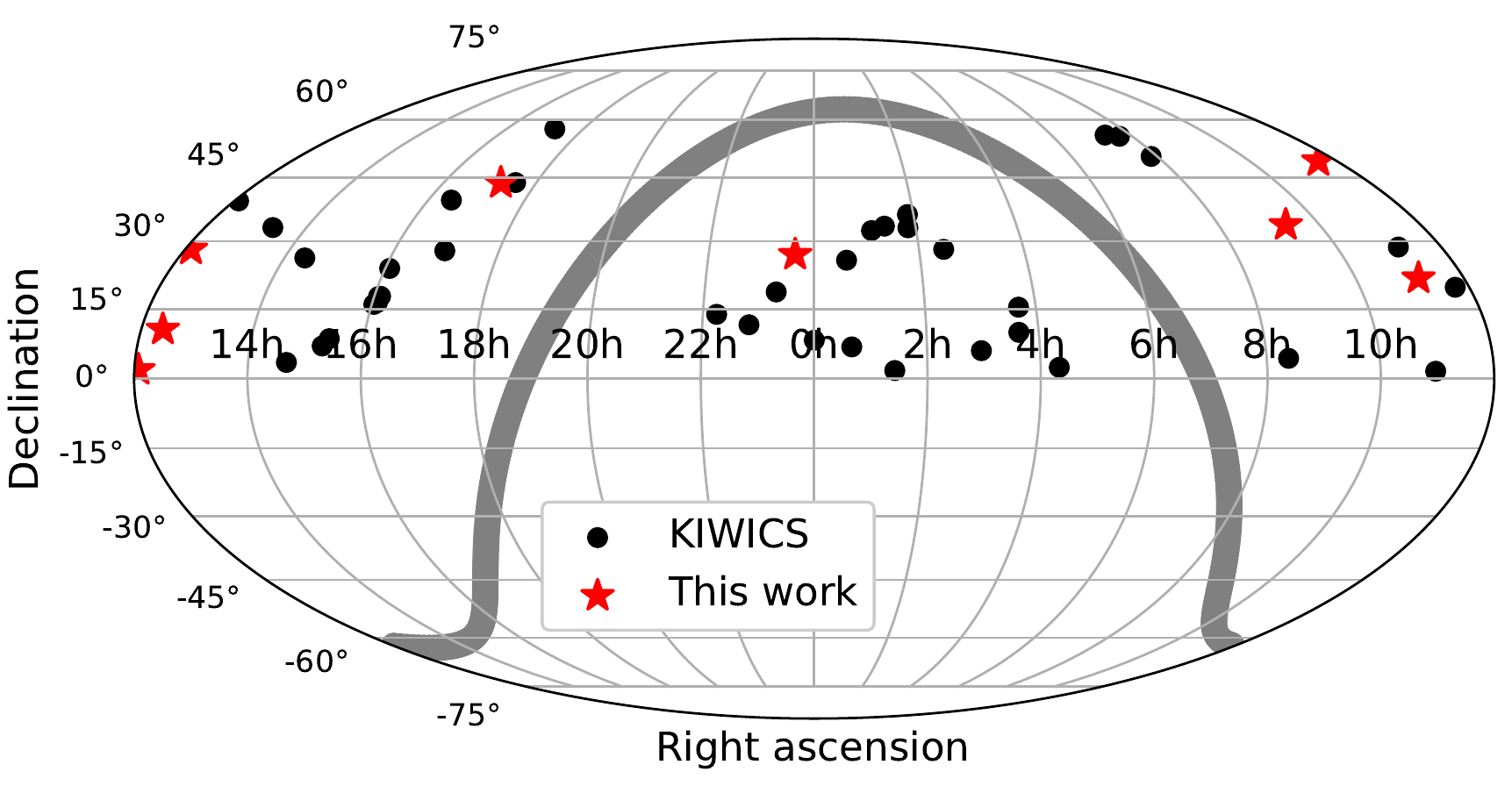}
\caption{Positions of the clusters being surveyed in \texttt{KIWICS} (black) and those studied in this work (red).}
\label{fig:clusters}
\end{figure}

\begin{figure}
\centering
\includegraphics[scale=0.475]{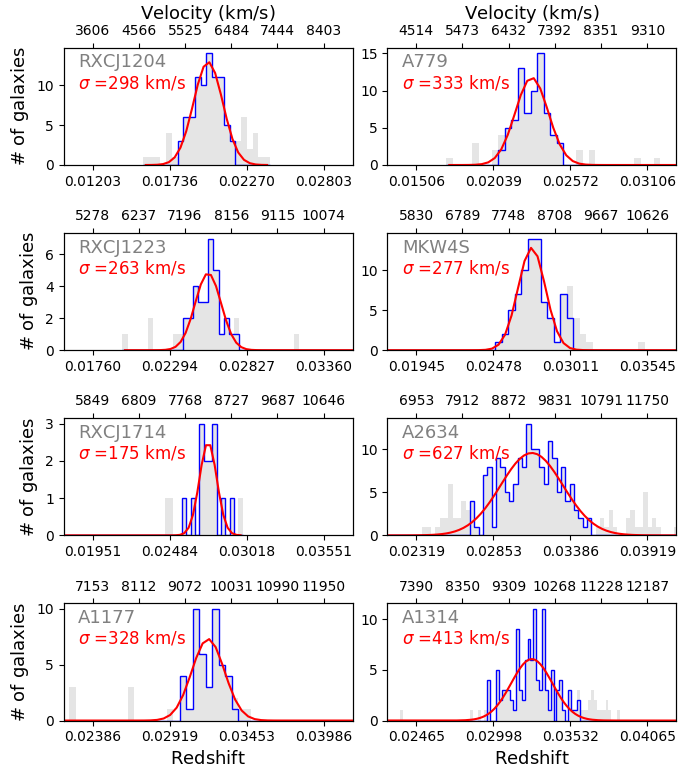}
\caption{Redshift distribution of the galaxies in our cluster sample, used to derive their redshift and velocity dispersion. grey histograms show the redshift distribution near our FOVs, the blue lines enclose the galaxies considered as part of the cluster (using a 2$\sigma$ criterion to select the main cluster structure), and the red lines the Gaussian fits.}
\label{fig:redshifts}
\end{figure}

\begin{figure*}
\centering
\includegraphics[scale=0.64]{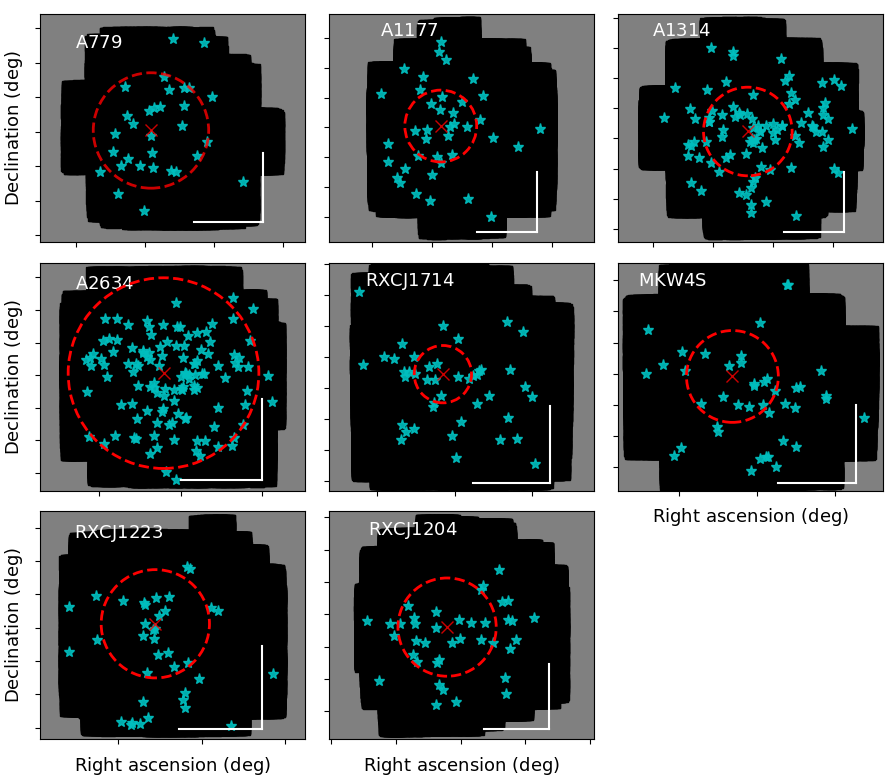}
\caption{FOVs (black area), R$_{200}$ (red dashed line) and hosted UDGs (aqua stars) of each cluster in our sample. The white rules show a scale of 0.5 deg.}
\label{fig:positions}
\end{figure*}

Figure \ref{fig:clusters} shows a map of the sky with the clusters being surveyed and those studied in this work. As can be seen, our clusters are not particularly located near the Galactic disk, and we do not have a strong cirrus contamination in our images. Table \ref{tab:sample} presents our sample, giving the coordinates of each cluster, its redshift, M$_{200}$, R$_{200}$ and mean seeing during the observations. The coordinates point to the X-ray center of the cluster, according to \cite{piffaretti}, and the redshifts are derived by fitting a Gaussian function to the redshift distribution as given in SDSS and NED databases and taking its mean. Figure \ref{fig:redshifts} shows the redshift distributions of our clusters.
For getting M$_{200}$ we follow \cite{munari}, deriving a velocity dispersion from the same Gaussian fit, and correcting for cosmological expansion by $\sigma$ = $\sigma_{obs}/(1+z)$. Then, R$_{200}$ is derived from M$_{200}$ assuming spherical symmetry and a density equal to 200 times the critical density of the Universe at the redshift of each cluster. The uncertainties in our mass determinations come from assuming an error of 10\% in the estimation of $\sigma$ and propagating the errors. 10\% was found to be a robust uncertainty for $\sigma$ by testing the dispersion in the values of $\sigma$ when changing the width of the bins of the redshift distribution as well as removing points of the distribution. Figure \ref{fig:positions} shows a schematic figure of our FOVs and the circle subtended by each R$_{200}$, illustrating also our spatial coverage and the UDGs found in this work.

	\begin{table*}
	\caption{Name, coordinates, redshift, M$_{200}$, R$_{200}$ and mean seeing during the observations for the eight clusters in our sample.}
	\label{tab:sample}
	\begin{center}
	\begin{tabular}{lccccccc}
	\hline
	\noalign{\smallskip}
    Cluster & RA (J2000) & DEC (J2000) & Redshift & M$_{200}$ & R$_{200}$ & Seeing $r$-band \\
      & (hh:mm:ss) & ($^{\textnormal{o}}~:~^{'}~:~^{''}$) & &  ($\times$10$^{13}$M$_\odot$)& (kpc) & (arcsec)\\ \hline
    \noalign{\smallskip}
     RXCJ1204.4+0154   & 12:04:25.2  & +01:54:02 & 0.0200 & 2.9  $\pm$ 0.9 & 630  $\pm$ 60 & 1.51 \\ \noalign{\smallskip}
        Abell 779    &  09:19:49.2 & +33:45:37 & 0.0231 & 4.0  $\pm$ 1.2 & 700  $\pm$ 70 & 1.41 \\ \noalign{\smallskip}
     RXCJ1223.1+1037   & 12:23:06.5  & +10:27:26 & 0.0256 & 2.0  $\pm$ 0.6 & 550  $\pm$ 60 & 1.63 \\ \noalign{\smallskip}  
        MKW4S   & 12:06:37.4  & +28:11:01 & 0.0274 & 2.3  $\pm$ 0.7 & 580  $\pm$ 60 & 1.43 \\ \noalign{\smallskip}
     RXCJ1714.3+4341   & 17:14:18.6  & +43:41:23 & 0.0275 & 0.6  $\pm$ 0.2 & 370  $\pm$ 40 & 1.32 \\ \noalign{\smallskip}
		Abell 2634   & 23:38:25.7  & +27:00:45 & 0.0312 & 26.6 $\pm$ 8.0 & 1310 $\pm$ 130 & 1.55 \\ \noalign{\smallskip}
        Abell 1177   & 11:09:43.1  & +21:45:43 & 0.0319 & 3.8  $\pm$ 1.1 & 690  $\pm$ 70 & 1.51 \\ \noalign{\smallskip}
        Abell 1314   & 11:34:50.5  & +49:03:28 & 0.0327 & 7.6  $\pm$ 2.3 & 870  $\pm$ 90 & 1.54 \\ \noalign{\smallskip}
      \hline
	\end{tabular}
	\end{center}
	\end{table*}

\section{Detection of the UDGs}
\label{sec:detection}
\subsection{Definition of UDG in this work}
\cite{vandokkum} defined a UDG as a galaxy with $\mu(g,0)$ > 24 mag arcsec$^{-2}$ and $R_e$ > 1.5 kpc. However, there is no a physical motivation to choose those particular values (see RT17b), and this has lead to different definitions of a UDG through the literature. 
For instance, \cite{roman} considered galaxies with $\mu$(g,0) > 24.0 mag arcsec$^{-2}$ (as measured via S\'ersic profile fitting), RT17b used $\mu$(g,0) > 23.5 mag arcsec$^{-2}$, \cite{Aku} $\mu(r^\prime,0)$ > 23.0 mag arcsec$^{-2}$, while \cite{koda} and vdB+16 used the mean effective surface brightness of $\langle\mu(R,R_e)\rangle$ > 24 mag arcsec$^{-2}$ and $\langle\mu(r,R_e)\rangle$ > 24.0 mag arcsec$^{-2}$, respectively. As the last authors explain, the mean effective surface brightness within the effective radius, $\langle\mu(R_e)\rangle$, is more related to the detectability of galaxies than the central surface brightness and has also the advantage that, given a fixed surface brightness and effective radius, it is independent of the S\'ersic index.

These properties of the mean effective surface brightness, coupled with the fact in Paper I we were interested in comparing our results with vdB+16, motivated us to also work with the quantity $\langle\mu(r,Re)\rangle$, and for our analysis here we adopt the definition of a UDG being a galaxy with effective radius\footnote{As for the surface brightness, there is no consensus in the literature, and while some authors use the effective radius as the semi-major axis length of an ellipse fitted to the galaxy isophotes enclosing half of the galaxy's total light, $R_e$, others prefer the ``circularized" effective radius $R_{e,c}$ = $R_e\cdot \sqrt{b/a}$. We use the (non-circularized) effective radius.} $R_e$ $\geq$ 1.5 kpc and $\langle\mu(r,R_e)\rangle \geq$ 24 mag arcsec$^{-2}$. We also demand a S\'ersic index $n$ < 4 and a color $g-r$ < 1.2 mag, with the aim of preventing contamination from concentrated and background objects, and colors not representative of stellar populations of low-$z$ galaxies\footnote{The choice of the limiting color $g-r$ < 1.2 comes from assuming a standard limiting color like $g-r$ < 0.8-1.0 (e.g. \citealt{agulli}; \citealt{Aku}) plus giving some extra freedom to the color, to be on the safe side considering uncertainties.}. The constraint on the S\'ersic index is relatively weak and allows including relatively concentrated objects. We keep this value for consistency with the literature, but we notice that our sample is not biased to high-$n$ galaxies: only <3\% of the UDG candidates reported here have $n > 2$.

\subsection{Selection and characterization of UDG candidates}
We use \texttt{SExtractor} (\citealt{sextractor}; \citealt{sexford}) to identify potential UDG candidates and then \texttt{GALFIT} \citep{galfit} to derive more accurate photometry via a fit to their light profiles. However, before using \texttt{SExtractor} in our images, we perform a series of simulations which allow us to: i) retrieve a high recovery fraction, ii) determine the most efficient way to run \texttt{SExtractor} without detecting many false positives, which is of importance given our large sample, iii) estimate the completeness levels in our images, relevant when comparing with the literature and in between clusters, and iv) calibrate the output from \texttt{SExtractor}, essential when doing the selection of potential UDGs.

\begin{figure*}
\centering
\includegraphics[scale=0.39]{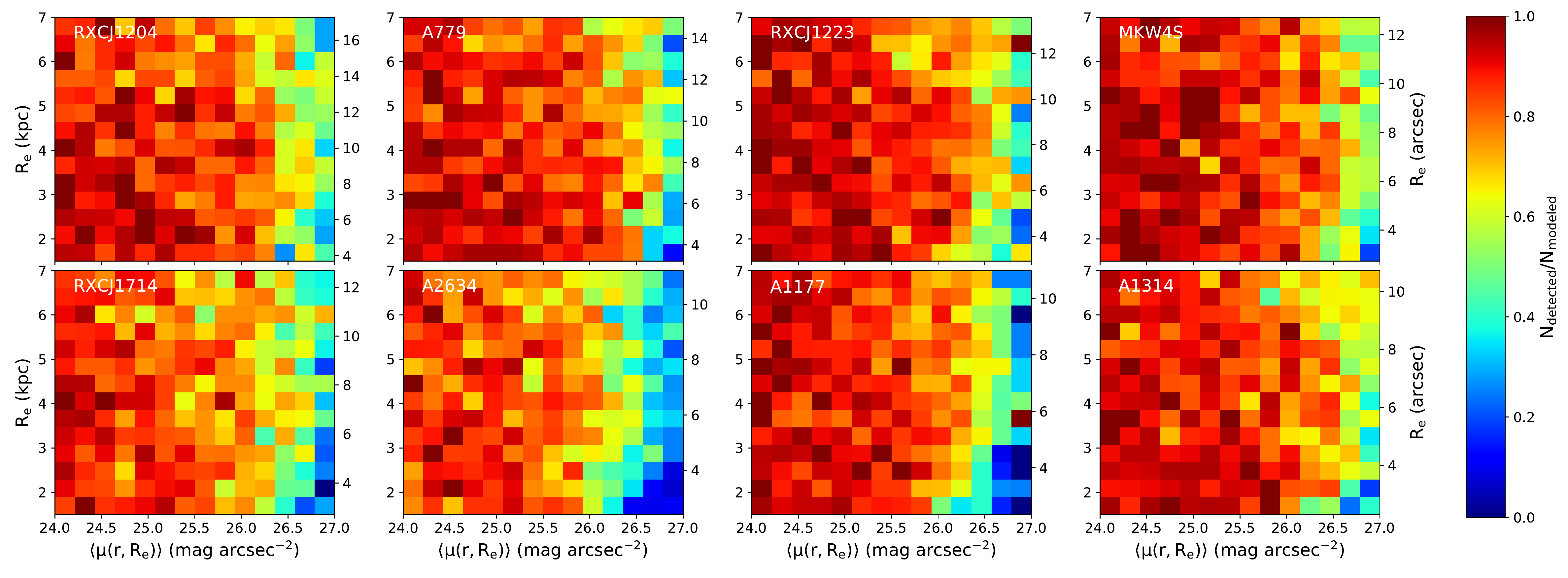}
\caption{Expected completeness levels for the UDGs in each FOV (indicated by the white label). The figure shows the surface brightness--size plane, with the color showing the recovery fraction. The y-axis shows both angular and physical scales.}
\label{fig:completeness}
\end{figure*}

We simulate UDGs as follows. For each cluster, we model galaxies with the same parameter space as UDGs in effective radius (here we assume that the galaxies are at the distance of each cluster, and we work in physical units), surface brightness, S\'ersic index and axis ratio, using a 2-D S\'ersic profile. The galaxies are convolved with the PSF profile of the cluster (see Section \ref{subsec:galmod} below) and Poisson noise is applied to each pixel. Finally the galaxies are injected into each cluster image; in practice we simulate 5000 galaxies for each cluster, injecting only a fraction of them ($50-100$) each time, repeating the procedure several times. We then run \texttt{SExtractor}, varying especially the detection parameters of the detection threshold (\texttt{DETECT\_THRESH}) and the minimal detected area (\texttt{DETECT\_MINAREA}) until we are able to retrieve a high recovery fraction without too many false detections (to check this we also generated mock galaxies injected in a representative background for each cluster, with no additional sources injected, allowing us to test the number of false detections). Figure \ref{fig:completeness} shows the expected recovery fractions in the size-surface brightness plane; while there are a few differences between clusters the general behavior is the same and we do not expect these small differences to play a role against the homogeneity of our dataset. We note also the similarities with the completeness levels in vdB+16 (see their Figure 1, and notice also the slightly different color scheme). Figure \ref{fig:completeness} has been already corrected for the expected bias or difference between the data and \texttt{SExtractor} measurements; this bias is almost the same for all the clusters, being $\sim$ 0.3 kpc for the effective radii and $\sim$ 0.35 mag arcsec$^{-2}$ for the surface brightness. It is also worth to mention that to consider a mock galaxy as detected, a detection near its centroid was required, as well as a measure of its effective radius and surface brightness close to the actual mock values. Appendix \ref{sec:A2} enlists the non-default \texttt{SExtractor} parameters used for each cluster.

Once we know the best configuration for \texttt{SExtractor} from the simulations, we run it on the real cluster images in dual mode, with the $r$-band image used to detect the sources. The main parameters obtained from \texttt{SExtractor} are \texttt{ALPHA\_J2000, DELTA\_J2000, MAG\_AUTO, FLUX\_RADIUS, MU\_MEAN\_MODEL, FWHM\_IMAGE, FLAGS} and the \texttt{CLASS\_STAR} stellarity index\footnote{Within the \texttt{SExtractor} environment, \texttt{ALPHA\_J2000} and \texttt{DELTA\_J2000} refer to the coordinates of the center of the objects, \texttt{MAG\_AUTO} is the Kron-like magnitude, \texttt{FLUX\_RADIUS} is the radius containing some percentage of the total galaxy flux, \texttt{MU\_MEAN\_MODEL} is the mean effective surface brightness, \texttt{FWHM\_IMAGE} the FWHM for each source in the FOV, \texttt{FLAGS} is a label regarding how well the photometry was done according to \texttt{SExtractor} and the \texttt{CLASS\_STAR} parameter is a probability assigned to each source to be a star (\texttt{CLASS\_STAR} $\sim$ 1) or a galaxy (\texttt{CLASS\_STAR} $\sim$ 0).}. We keep the objects with \texttt{CLASS\_STAR} $\leq$ 0.2 and \texttt{FLAGS} < 4, to select objects that are highly likely to be galaxies and with relatively good photometry.

We use this preliminary photometry to select the potential UDGs, based on their surface brightnesses and effective radii (we use the \texttt{FLUX\_RADIUS} containing 50\% of the galaxy light as a proxy of the effective radius).
To set the limits of our selection box we considered the mentioned bias in effective radius and surface brightness, i.e. we set the selection border as the lower limits of our definition of UDGs minus the expected biases between the real data and the photometry of \texttt{SExtractor}.

We realize from the simulations that we will miss a number of UDGs, either because they are not detected since they are too faint or because their photometry is not good enough, making them to lie outside our searching region. Notwithstanding, the bulk of the UDGs should be detected, and in any case, this is a latent problem in the automatic detection techniques; meaning that the fraction of lost UDGs should be similar to the fraction of lost UDGs in the literature.

As a final step of our selection of potential UDGs, and with the aim of ensuring the purity of our sample, we do a visual cleaning of the candidates. We perform a visual inspection of individual stamps of each galaxy, and we remove artefacts if present (e.g. tidal structures near galaxies, multiple detections in haloes of bright galaxies or in the spikes of saturated stars or multiple small objects not separated by \texttt{SExtractor}). The remaining galaxies go to the next step, where we characterize them with \texttt{GALFIT}.

\subsection{\texttt{GALFIT} modeling}
\label{subsec:galmod}
To characterize our galaxies we use \texttt{GALFIT} to fit a S\'ersic \citep{sersic} profile of the form:
\begin{equation}
\Sigma(r) = \Sigma_e \cdot e^{-k\left[(r/R_e)^{1/n} -1\right]} ,
\end{equation}
\noindent
where $\Sigma_e$ is the surface brightness (in flux units) of the galaxy at its effective radius $R_e$, n the S\'ersic index, and $k$ an index coupled with $n$.

To perform the fitting in a semi-automatic mode, we use the pipeline and general procedure from \cite{Aku2} (but see also \citealt{Aku}). A complete description is given in those references, so here we just briefly summarize the main steps followed for each cluster:

Stamps of each potential UDG are generated, as well as its $\sigma$-image. The size of each cutout is 10 times the \texttt{SExtractor} effective radius. We mask the sources near each galaxy we want to fit, to account only for the light component we are interested in; as in \cite{Aku} and \cite{Aku2} initial masks are generated with \text{SExtractor} by masking all the sources larger than 100 pixels and brighter than 1$\sigma$ and then removing all the masks closer to 2 times the effective radius of the galaxy. By stacking bright non-saturated stars from each image, we build a PSF profile in both bands, to be used later for \texttt{GALFIT}.

To derive a proper initial seed for \texttt{GALFIT} we make a radial profile of the galaxy using circular bins with width of two pixels, and taking then the average of each bin to make a cumulative profile extending 3 effective radius (as measured by \texttt{SExtractor}). With this we build the growth curve and its effective radius and magnitude are given to \texttt{GALFIT}. We let \texttt{GALFIT} fit the S\'ersic profile freely\footnote{Some studies have studied the fraction of nucleated UDGs (e.g. \citealt{koda, Aku}), but the resolution of our data does not allow us to observe nucleation. Also, some works constrained the S\'ersic index when doing the \texttt{GALFIT} fitting (e.g. vdB+16, to be higher than 0.5), but we do not set constraints in the S\'ersic index.} in both bands, but keep the $r$-band derived parameters, since the S/N is higher for that filter and the $r$-band is less affected by the light of young stars.

A model is considered good if i) it visually resembles the real galaxy, ii) its radial profile mimics the observed radial profile, and iii) the residuals are low and the fit produces a good $\chi ^2 _\nu$ parameter. Bad fits appear if the masking is not good or if the fitting areas are too small; in those cases we run \texttt{GALFIT} again with a slightly different configuration, until a robust result is reached.  All our models converged. If the final residuals (looking at the image-model images in magnitude units) show strong substructures not typical of UDGs (e.g. features characteristics of large spirals like prominent bulges, bars or continuous spiral structures) or seem to be background galaxies close to each other, we reject them. A caveat to be considered is the subjectiveness of this analysis, however each case was carefully checked multiple times trying to reduce any bias. In any case the fraction of rejected galaxies was always small.

From \texttt{GALFIT} we get the center of each galaxy, its effective radius, axis ratio, S\'ersic index, and position angle. With these parameters we are able to measure the magnitudes at the effective apertures, and thus the color. Hereafter we work with magnitudes and colors measured at the effective aperture (i.e. we performed aperture photometry using the effective radius, axis ratio and position angle obtained from \texttt{GALFIT}, see also \citealt{Aku2}).
This color is more stable than using the total color because, while they are not very different for most of the galaxies (the mean difference is 0.04 mag), it reduces the error by a factor 2-3, because it is less dependent on systematic errors in the sky background determination. The mean effective surface brightness is derived from:
\begin{equation}
\langle\mu(r,R_e)\rangle = m(<R_e) + 2.5\log(\pi (b/a) R_e ^2)~.
\end{equation}
\noindent
Magnitudes and colors are corrected for Galactic extinction and $k-$corrections, taken from \citet{schlafly} and \citet{chilingarian}, respectively. Cosmological dimming \citep{tolman1,tolman2} is also considered (see for instance \citealt{reviewlsb}).

\begin{figure*}
\centering
\includegraphics[scale=0.67]{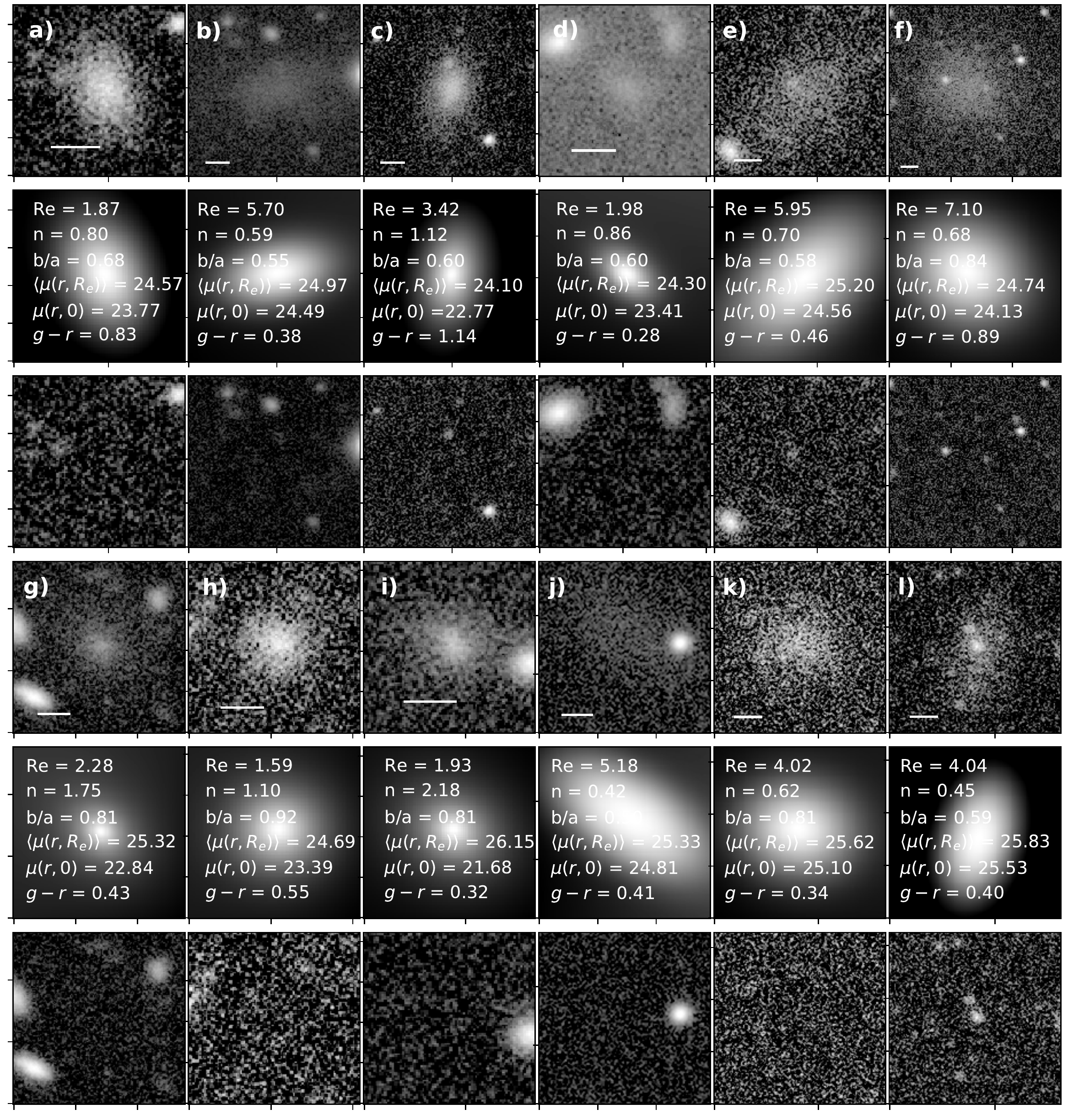}
\caption{Examples of UDGs found in different clusters. Top panels show the $r$-band image of each UDG, middle panels the GALFIT models with their structural parameters, and bottom panels the residuals of the fits. The color scale is logarithmic to highlight the low-surface-brightness structures and the white bands in the top panels show a scale of 5 arcsec. The figure has been taken from Paper I.}
\label{fig:examples}
\end{figure*}

Finally, with all the structural parameters known, we find which galaxies fulfill our definition of a UDG (under the assumption that they lie at the redshift of their associated cluster), and they constitute our final sample. In the end we find 442 UDG candidates in our eight FOVs, with 247 of them lying at projected distances within 1 R$_{200}$. In the rest of this paper we analyze and discuss the implications of the structural parameters of these UDGs. Figure \ref{fig:examples} shows examples of galaxies with different parameters classified as UDGs (Paper I). A table containing the mean structural parameters of our sample is presented below, and the whole catalogue is available upon request.

The uncertainties associated with the color measurements are of the order 0.1-0.2 mag, and they are estimated as in \cite{Aku}:
\begin{equation}
\begin{split}
\sigma^2 _{g-r} = \sigma^2 _{ZP,g} + \sigma^2 _{ZP,r} + \left(\dfrac{2.5}{I_r \ln 10}\right)^{2} (\sigma_{I,g} + \sigma_{sky,g})^2 \\ 
+ \left(\dfrac{2.5}{I_g \ln 10}\right)^{2} (\sigma_{I,r} + \sigma_{sky,r})^2~ ,
\end{split}
\end{equation}
where $I_{g,r}$ is the mean intensity within the effective aperture in each band, $\sigma_{I,g,r}$ is its error, and $\sigma_{ZP,g,r}$ and $\sigma_{sky,g,r}$ are the errors in the zero-point and sky determination, respectively.

\begin{figure}
\centering
\includegraphics[scale=0.46]{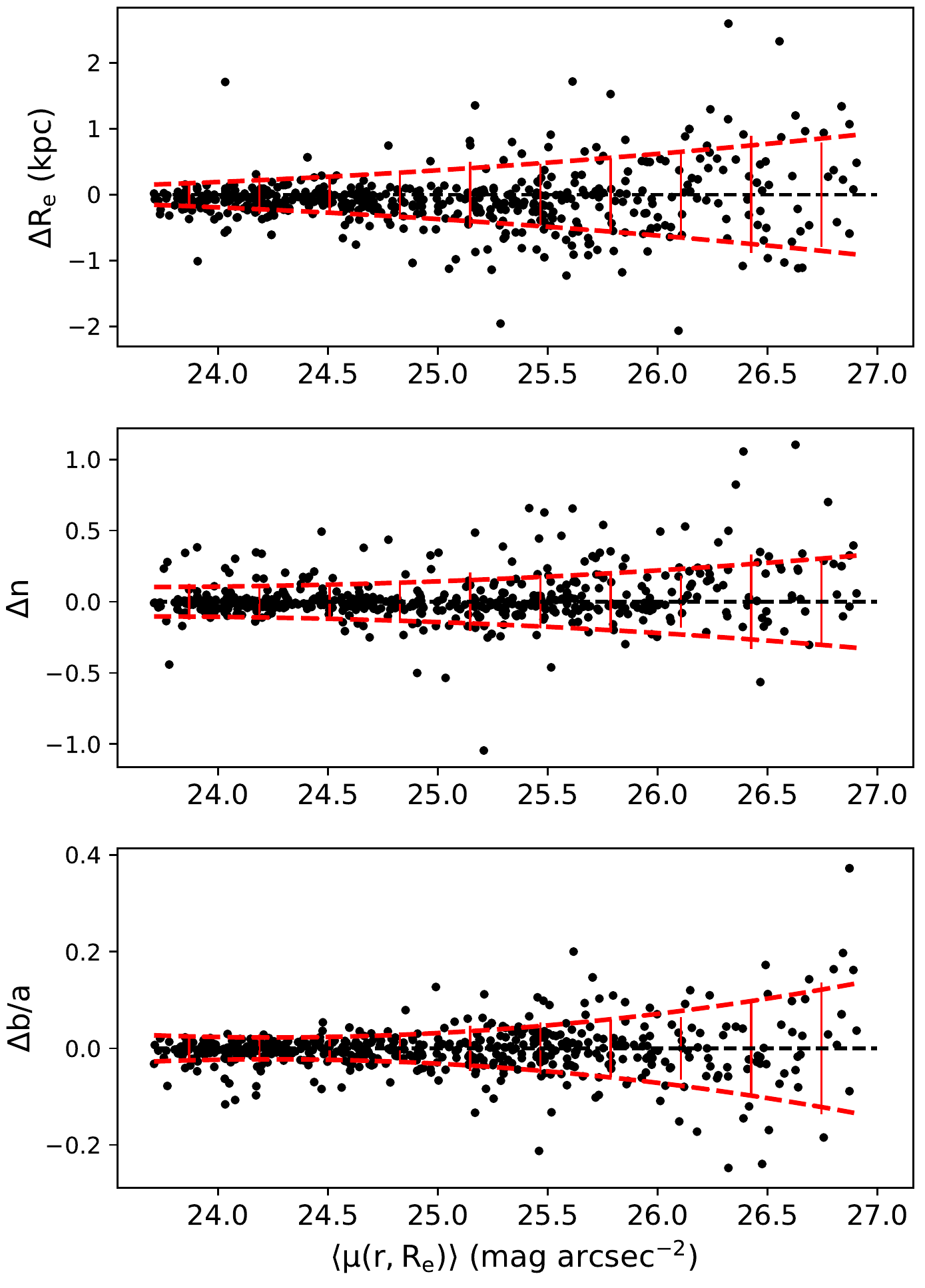}
\caption{Difference between the model and recovered parameters of mock galaxies, as a function of the surface brightness. $\Delta$ refers to model-GALFIT values and the black lines shows the line $\Delta$ = 0. Red solid vertical lines are the 2$\sigma$ errors in bins of surface brightness, while the red dashed lines show the second degree polynomial fit used to characterize our uncertainties.}
\label{fig:galfiterrors}
\end{figure}

To characterize the uncertainties in the effective radius, S\'ersic index and axis ratio, we use again mock galaxies to look at the differences between modeled and recovered parameters. 
This time we generate 500 galaxies as we did above: covering all the parameter space typical of UDGs, convolving in this case with the mean PSF profile of our sample and adding Poisson noise to each pixel. The result of the comparison is shown in Figure \ref{fig:galfiterrors}, where the differences in the parameters are plotted as a function of the surface brightness, since it dominates the uncertainties. We find 2$\sigma$-clipped mean offsets (model -- \texttt{GALFIT}, blue lines in Figure \ref{fig:galfiterrors}) for each parameter of $\bar{\Delta}R_e = -0.082$ kpc, $\bar{\Delta}n = -0.017$, and $\bar{\Delta}b/a = -0.003$.

To quantify the uncertainties, we  measure the standard deviation, at a 2$\sigma$ level, of each parameter for different bins of surface brightness (vertical red solid lines), and we fit to them a second degree polynomial of the form $\Delta$ = $a$x$^2$ $+ b$x $+c$ (red dotted lines). Table \ref{tab:coefserror} gives the values for the $a$, $b$ and $c$ coeficients for each quantity analyzed. For illustration, the uncertainties in the three parameters at $\langle \mu(r,R_e) \rangle$ = 26 mag arcsec$^{-2}$ are $\delta R_e \sim \pm$ 0.60 kpc, $\delta n \sim \pm$ 0.23 and $\delta b/a \sim$ 0.07. 

\begin{table}
\caption{Parameters of the polynomials of degree 2 used for deriving the uncertainties associated with each quantity fitted with \texttt{GALFIT}.}
\label{tab:coefserror}
\begin{center}
\begin{tabular}{lccc}
	\hline
	\noalign{\smallskip}
        Parameter          & $a$     &$ b$         & $c$ \\ \hline    \noalign{\smallskip} 
        $\Delta R_e$  & 0.03562 & -1.56724 &17.28818 \\  \noalign{\smallskip}
        $\Delta n $ & 0.02003 & -0.94497 &11.24596 \\  \noalign{\smallskip}
        $\Delta b/a $& 0.01577 & -0.76446 & 9.28807 \\  
    \hline
\end{tabular} 
\end{center}
\end{table}

\section{Structural parameters}
\label{sec:params}
\begin{figure*} 
\centering
\includegraphics[scale=0.6]{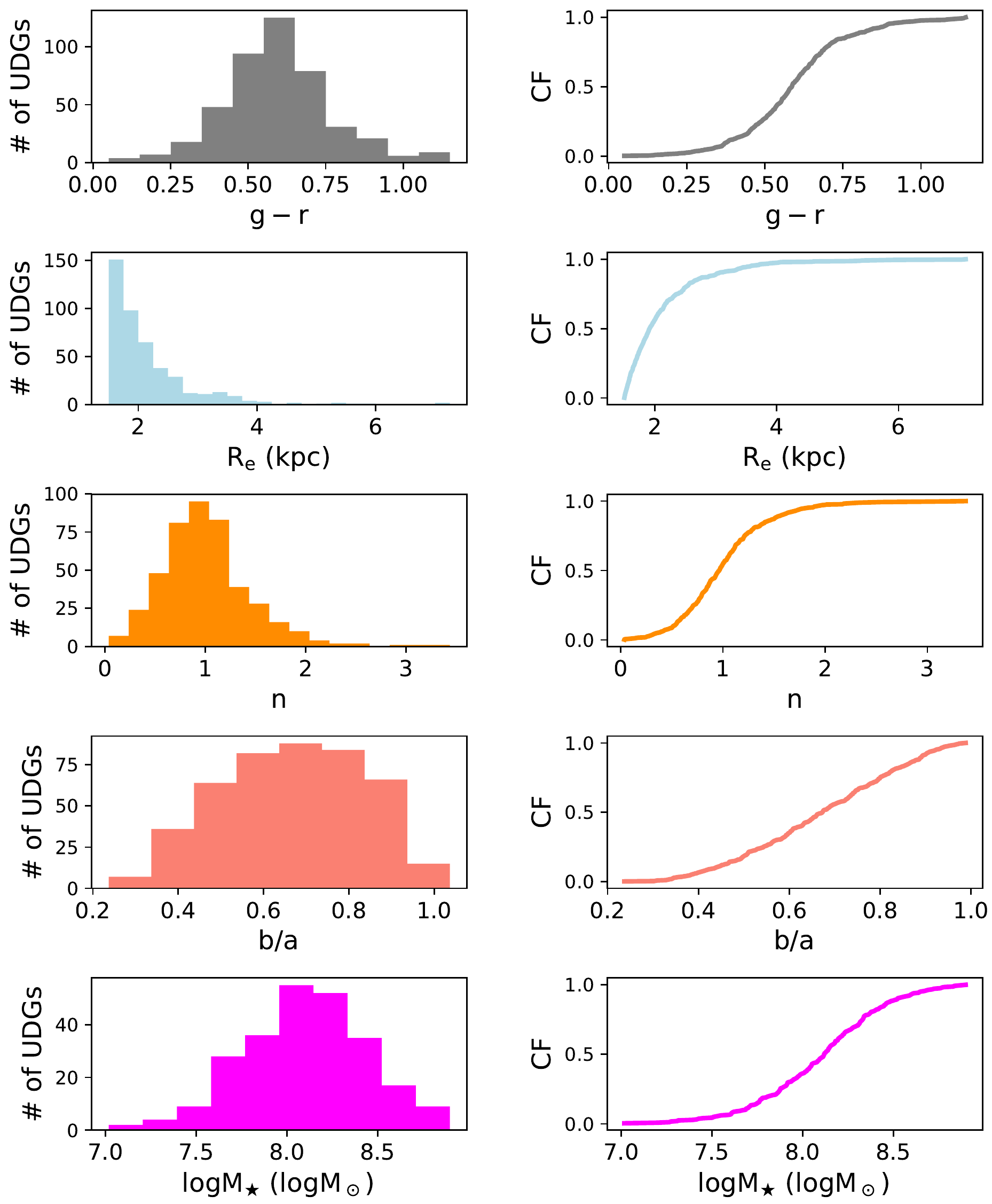}
\caption{Histograms (left) and cumulative fractions (right) of the distributions of structural parameters of the 442 UDG candidates found in this work. See text for details. The effective radius distribution assumes that the galaxies lie at the redshift of their associated cluster.}
\label{fig:histogram}
\end{figure*}

Figure \ref{fig:histogram} shows the distributions of colors, effective radii, S\'ersic indices and axis ratios of our UDGs sample. The histogram and cumulative fraction of the color show, as expected, a passively evolving population, although UDGs show a range in color\footnote{A caveat regarding a selection bias should be taken into account. As discussed in \cite{trujillo} and Paper I, blue UDG-like galaxies are brighter than the red ones, which makes them escape the surface brightness criteria usually used in the literature. Therefore we are bias to find more red UDGs than bluer, brighter, counterparts.}. The effective radius distribution is highly dominated by UDGs smaller than 2 kpc, as observed in other clusters (e.g. vdB+16, \citealt{roman}). The surface brightness profiles are very close to exponential (note also the almost nonexistent population of UDGs with $n > 2$, only 3\% of our sample), and the (apparent) axis ratio distribution seems to resemble the expected distribution of thick disks. As a population our UDGs have median (mean) values of $g-r$ = 0.59 (0.59), $R_e$ = 1.91 (2.16) kpc, $n$ = 0.96 (1.01), and $b/a$ = 0.67 (0.67). 

For completeness, we also derive rough stellar masses for the UDGs. This is done using mass-to-light ratio relations. Specifically, we use the relation by \cite{m2l}:
\begin{equation}
\log(M_\star/L)_\lambda = 1.629\times (g-r) - 0.792 ,
\end{equation}
assuming a Chabrier Initial Mass function \citep{chabrier}. The values we find are in good agreement with the literature (e.g. vdB+16, RT17b), having our sample a median (mean) stellar mass of 1.3 (2.0) $\times$10$^8$ M$_\odot$. 

We also find the structural parameters for the UDGs inside 1 (projected) R$_{200}$ for each cluster separately, as given in Table \ref{tab:params} in the Appendix \ref{sec:A1}, where the mean, median, minimum and maximum of each parameter is shown. The means of all the medians of each parameter (denoted by $\langle~\rangle$) are $\langle g-r \rangle$ = 0.61, $\langle R_e \rangle$ = 1.95 kpc, $\langle n \rangle$ = 1.00 and $\langle b/a \rangle$ = 0.72. Given the expected low contribution of interlopers in our sample (see for instance Paper I for details on the statistical background decontamination when studying the abundance of UDGs, as well as Section \ref{subsec:spatialdistribution}), and the fact that the structural parameters of these interlopers do not significantly differ from the parameters of the bulk of the UDG population, we do not attempt to correct the statistical distributions of our mean parameters, since this correction would not change them.

\subsection{Scaling relations of UDGs}
\label{sec:scalingrelations}
We now examine the photometric scaling relations of UDGs. We compare these relations with different types of galaxies: galaxies from the Fornax Deep Survey (FDS, \citealt{Aku2}) and a set of bright (M$_r$ < -18 mag) galaxies in low-$z$ clusters \citep{sanchez}. 
The parameters for these galaxies were originally derived by fitting S\'ersic profiles to their light distribution. In the case of the bright galaxies of \cite{sanchez}, we use the equations by \cite{bilir}\footnote{$g=V+0.634\times(B-V)-0.108$ and $g-r = 1.124\times(B-V)-0.252$.} to covert their photometry to our filters, and we covert their sizes and luminosities to our cosmology. For the FDS galaxies these corrections are not needed. For both datasets we apply $k-$ and surface brightness dimming corrections (they were already corrected by Galactic extinction), in the same way as for our data.

\begin{figure*}
\centering
\includegraphics[scale=0.6]{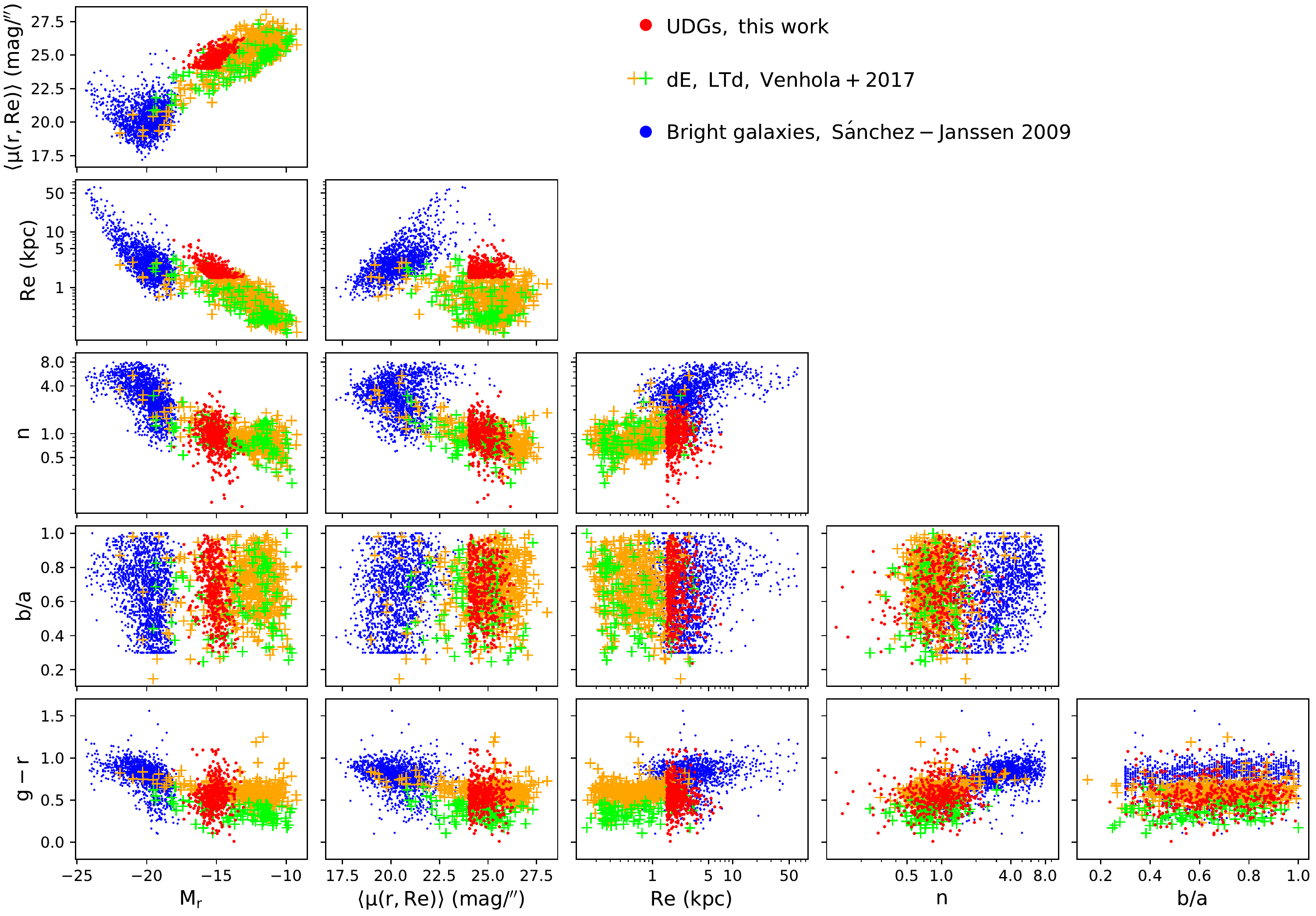}
\caption{Scaling relations of UDGs compared with other types of galaxies. Red points show our UDGs sample; orange and lime crosses are the FDS early- and late-type dwarfs, respectively; blue points represent bright galaxies of nearby clusters.}
\label{fig:scaling}
\end{figure*}

\begin{figure*}
\centering
\includegraphics[scale=0.48]{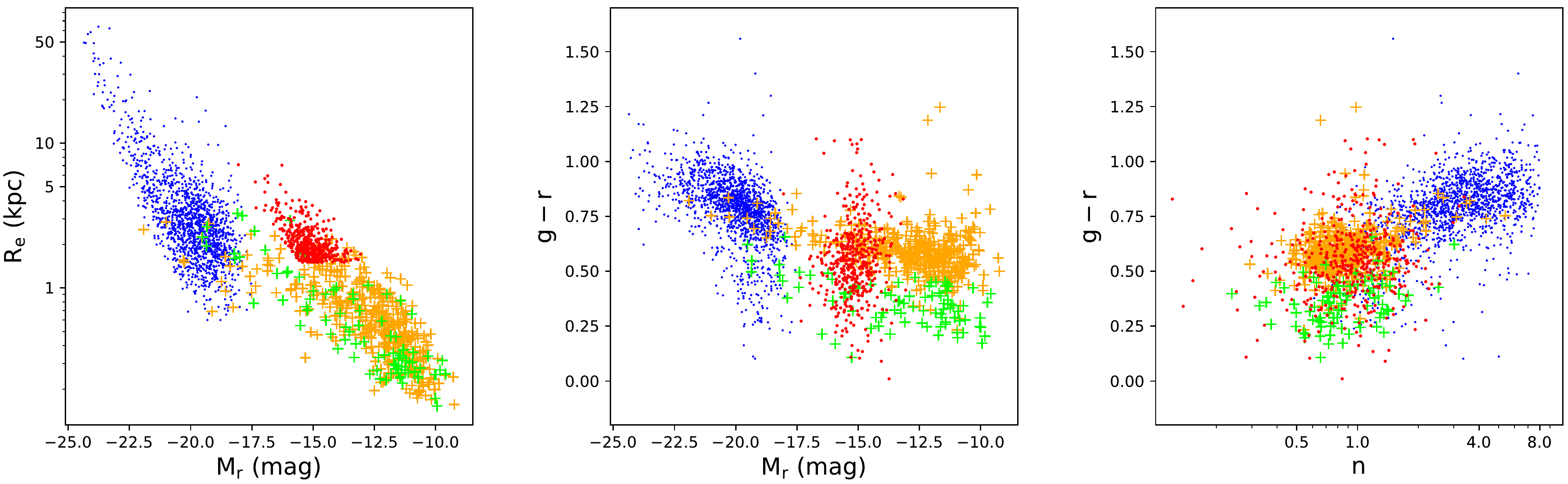}
  \caption{M$_r$--$R_e$, color--M$_r$, and n--color diagrams for our sample of UDGs compared with other types of galaxies. Symbols are as in Figure \ref{fig:scaling}.}
  \label{fig:zoomscaling}
\end{figure*}

The result of the comparison is shown in Figure \ref{fig:scaling} and Figure \ref{fig:zoomscaling}, where UDGs are plotted with red points, dEs and late-type dwarfs from the FDS with orange and lime crosses, respectively, and the bright galaxies from \cite{sanchez} with blue points.

The general picture seems to be clear, with UDGs being a part of a continuous distribution between dwarf and giant galaxies, as previously discussed in the literature (e.g. \citealt{Aku}; \citealt{conselice}, and references therein). Apart from being the bridge between small and large galaxies, UDGs behave very similar to other dwarfs, just standing out (by definition) for being the tail of the size distribution.

UDGs fit also very well on the color-magnitude diagram, with typical red sequence (RS) colors, although a few of them have colors bluer than the RS, which may be due to ongoing or recent star formation. We note that late-type and early-type dwarfs in the FDS are clearly separated in two sequences, however, when the sample of UDGs is included they populate a cloud in the color-magnitude diagram. While the uncertainties in our colors are of the order $\sim$ 0.2, it is likely that the color-magnitude relation is suggesting that UDGs are a mix of galaxies with both morphologies (cf. \citealt{sandage}). The same phenomenon is also observed in other scaling relations.
Additionally, we can see that UDGs follow roughly, with a larger scatter, the same color-S\'ersic index relation as other galaxies, with relatively high-$n$ galaxies showing redder stellar populations as they go from disk-like to more elliptical-like structures. 
In general, the photometric parameter space gives no sign of UDGs sharing common regions with galaxies with massive haloes, but of course this is not unexpected and the ultimate comparison should be done in the future with accurate dynamical mass determinations.

We will now focus on the distribution of UDGs in two specific scaling relations that may give hints to their origin.

\subsection{The $b/a-R_e$ plane}
\label{sec:baReplane}

It has been proposed by models that for a fixed stellar mass, larger UDGs should be more disk-like if they are high-angular momentum dwarfs (see for instance \citealt{dalcantondisks}). This can happen either if i) the dark matter haloes in which UDGs live have a high spin \citep{amorisco}, or if ii) UDGs are outliers in the angular momentum-mass relation because they have retained a higher-than-average fraction of the halo angular momentum \citep{posti}. In the latter case, UDGs do not necessarily inhabit high-spin haloes and their large sizes could still be possibly related to their SFHs \citep{dicintio}, for instance.
The observable signal of both these models is that at fixed mass, UDGs with larger $R_e$ have smaller $b/a$. Motivated by the idea above, \cite{Aku} compared the $b/a - R_e$ plane for Coma and Fornax cluster UDGs, finding that large UDGs in Fornax are more elongated than the smaller UDGs (a phenomena not observed in Coma) and reported a good agreement with the \cite{amorisco} model.



\begin{figure*}
\centering
\includegraphics[scale=0.55]{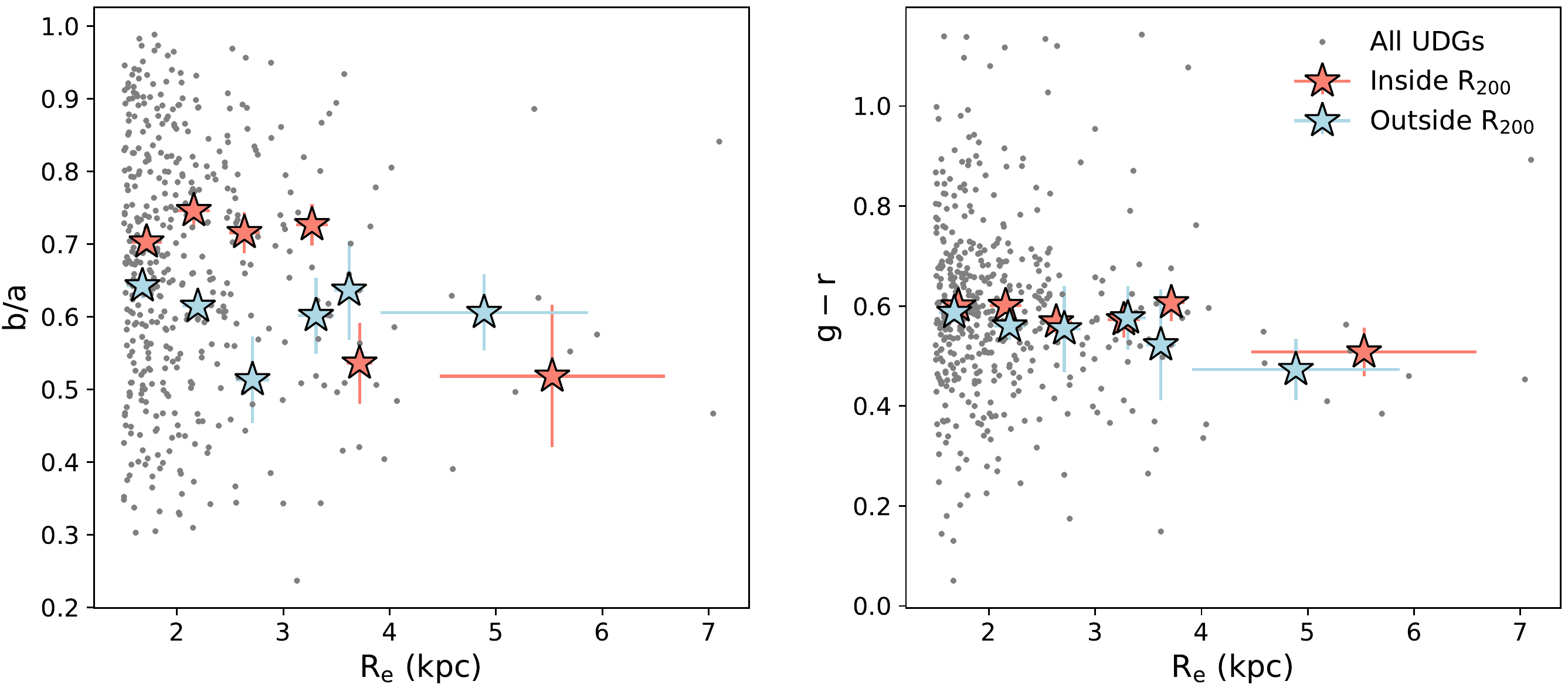}
\caption{\textit{Left:} $b/a - R_e$ plane for the UDGs in our sample. The grey points show all the UDGs in our sample, while the red and blue stars represent those galaxies inside and outside R$_{200}$, respectively. The distribution of the outer UDGs is flatter than the inner distribution, which is more concentrated towards higher axis ratios. The K-S comparison between the inner and outer distributions indicates that the distributions are statistically different. \textit{Right:} Color$- R_e$ plane. The symbols and colors are as in the previous figure. The distributions of inner and outer UDGs do not look different, and there is no strong correlation with the size of the galaxies and the colors. }
\label{fig:ARCOLvsRe}
\end{figure*}

In Figure \ref{fig:ARCOLvsRe} we show (left panel) the $b/a - R_e$ plane for our UDGs. It can be seen that no clear nor strong trends are visible for the whole population of UDGs (grey points).
Given that \cite{amorisco} considered UDGs in cluster environments, we show this relation in two different regimes: the high-density inner 1 R$_{200}$, and the relatively isolated low-density regions outside R$_{200}$\footnote{While we cover the inner R$_{200}$ for all the clusters, the coverage of the outer regions is not homogeneous, since the observations do not cover the same areas (relative to R$_{200}$) for each cluster, as seen in Figure \ref{fig:positions}.}. The stars representing both distributions show the medians of bins in effective radius and their standard errors.
We can see that while outside clusters UDGs of different sizes have the same axis ratios, inside clusters small UDGs are rounder than large UDGs. Moreover, the distribution of the axis ratios of the ``inner" UDGs is more concentrated towards higher axis ratios than for the ``outer" ones, for which the distribution becomes more flat or disky and shifts towards slightly lower axis ratios. The Kolmogorov-Smirnov (K-S) test on both axis ratio distributions confirms that the difference is statistically significant, with a p-value of 0.0002. This is a telltale clue of the environment affecting the axis ratio distribution of UDGs, and we will discuss this later. 

Regarding the predictions of the models discussed above, while for large UDGs the lower region on the panels are on average more populated than the upper region, UDGs in our clusters do not show a behavior as clear as in \cite{Aku} (although those authors found the UDGs in Fornax by visual inspection, and the largest UDGs are usually hard to identify with automatic softwares like \texttt{SExtractor}). We also split our sample in bins of stellar mass but the behavior of the binned data remains the same, finding with no trends. Caveats to take into account are that we are looking at apparent axis ratios and not to the intrinsic distribution, and that our estimation of the stellar mass is not extremely accurate.

Finally, it is interesting to notice that in the less-environmentally affected outer regions large UDGs are somewhat less elongated than large UDGs in the inner R$_{200}$. While it is important to keep in mind the scatter of our data and the fact that the axis ratios studied here are only the projected axis ratios, this may be an indication that originally large cluster UDGs are not necessarily more elongated than the smaller ones, but that they become very elongated only after they interact with the cluster environment, but more data are needed to draw conclusions from this $b/a - R_e$ plane.

\subsection{The color$- R_e$ plane}
\label{sec:colorReplane}
The models by \cite{dicintio} predict a correlation between size and SFHs (and gas fraction) of UDGs in isolation. Under the assumptions that galaxies with long or bursty SFHs have on average younger stellar populations than coeval galaxies with shorter SFHs, that gas-rich galaxies have younger stellar populations than gas-poor ones, and that color is a reliable tracer of the stellar population on UDGs, a relation between the size and the color of isolated UDGs would be expected. In a cluster environment galaxies are expected to be quenched and gas-deficient because their interactions with the cluster environment, but it is still interesting to look for signs of that phenomena in our clusters, especially in the outer (relatively more isolated) parts. On the other hand, the model by \cite{carleton} predict that larger cluster UDGs host older stellar populations (redder colors) than smaller UDGs. 

The right panel of Figure \ref{fig:ARCOLvsRe} shows the color$-R_e$ plane. Again, the grey points show all the UDGs while the stars show the separate groups of inner and outer UDGs. The general behavior of the color distribution is the same for both groups of UDGs. This lack of a relation between size and color of UDGs seems to be in disagreement with the predictions by \cite{carleton}. In principle our data point towards a scenario where the sizes of UDGs do not depend on their colors (and under some assumptions on their stellar populations and SFHs), but it is hard to compare with the model by \cite{dicintio}, since their UDGs are modeled in isolation, and the colors of our UDGs might be not tracing their original SFHs.

\section{The evolution of UDGs in clusters}
\label{sec:evolution}
We now focus on the evolution of UDGs in clusters. We study here three main aspects: the spatial distribution of UDGs, and the dependences of their structural parameters on the projected clustercentric distances and the masses of their host clusters. The abundance of UDGs in our sample has already been studied in Paper I, so we only comment briefly on it.

\subsection{A brief comment on the abundance of UDGs}

The abundance of UDGs (e.g. vdB+16, RT17b, Paper I) is interesting to study the evolution of UDGs. Table \ref{tab:udgs} gives the number of UDGs found in each cluster, and the only difference with the table presented in Paper I is that here we include all the galaxies with $R_e >$ 1.5 kpc, and not only those with $R_{e,c} >$ 1.5 kpc. For the eight \texttt{KIWICS} clusters studied here, under our definition of UDG, we find N(UDGs)$\propto$ M$_{200} ^{0.81 \pm 0.17}$; a sublinear slope at the 1$\sigma$ level. 

	\begin{table}
	\caption{The number of UDGs within R$_{200}$ for each cluster. The second and third columns give the number of UDGs found and the number corrected by background subtraction (BS), respectively. The fourth and fifth columns are the same as the second and third columns but consider only UDGs with circularized effective radii $R_{e,c} \geq$ 1.5 kpc.}
   \label{tab:udgs}
	\begin{center}
	\begin{tabular}{lcccc}
	\hline
	\noalign{\smallskip}
    Cluster & \multicolumn{2}{c}{N(UDGs)} &\multicolumn{2}{c}{N(UDGs, $R_{e,c}$)} \\
           &   raw & BS            &raw & BS\\ \hline
    \noalign{\smallskip}
     A779  & 24 & 22 &  21 & 20\\ \noalign{\smallskip}
    A1177  & 14 & 9  &   9 &  8 \\ \noalign{\smallskip}
    A1314  & 36 & 27 &  19 & 16\\ \noalign{\smallskip}
	A2634  &112 & 94 &  60 & 55\\ \noalign{\smallskip}
 RXCJ1714  &  8 &  7 &   7 &  7 \\ \noalign{\smallskip}
    MKW4S  & 14 & 11 &   5 &  5\\ \noalign{\smallskip}
 RXCJ1223  & 17 & 15 &  11 & 11\\ \noalign{\smallskip}  
 RXCJ1204  & 22 & 21 &  15 & 14\\ \noalign{\smallskip}
      \hline
	\end{tabular}
	\end{center}
	\end{table}

\subsection{Spatial distribution: galaxy alignments and radial surface density profile}
\label{subsec:spatialdistribution}

Our full coverage up to $\gtrsim$ 1 R$_{200}$ allows us to study the spatial location of UDGs in our eight galaxy clusters. This may encode information about the role of the environment shaping UDGs. \cite{yagi} reported an interesting feature in the spatial distribution of UDGs in Coma: an alignment between the major axis of the galaxies and the centre of the cluster. However, such behavior is not observed in the Fornax cluster \citep{Aku}, and it is interesting to check if it is present in our data.
To investigate this we compute the relative angles $\varphi$ between the major axis of our UDGs and the centre of each galaxy cluster, and this is compared with a flat distribution. To make sure our estimate of the position angle is accurate, we keep only UDGs with $b/a < 0.85$, as in \cite{yagi}, and inside 1 R$_{200}$. For all our clusters, a K-S test (comparing with the flat distribution) only rejects the null hypothesis (i.e. it neglects the probability of the distribution to be compatible with being flat) with a 95\% confidence level for the cluster Abell 1314, which shows radial alignment with an overabundance of galaxies with relative angles to the cluster centre of $\varphi \lesssim$ 20 deg. 30 UDGs in Abell 1314 met our selection criteria and were used for this analysis, so it is not likely that the alignment reported here is driven by randomness of low number statistics. As discussed in \cite{yagi}, mechanisms such as primordial alignment and tidal torques could explain galaxy alignments in clusters, but the analysis of this is hard to do with the current data and it is out of the scope of this work.

In general, the fact that only one out of our eight clusters show radial alignment of their UDGs suggests that this phenomena is not very common, and the orientation of most UDGs in clusters is not strongly affected by the environment.\\



\begin{figure}
\centering
\includegraphics[scale=0.4]{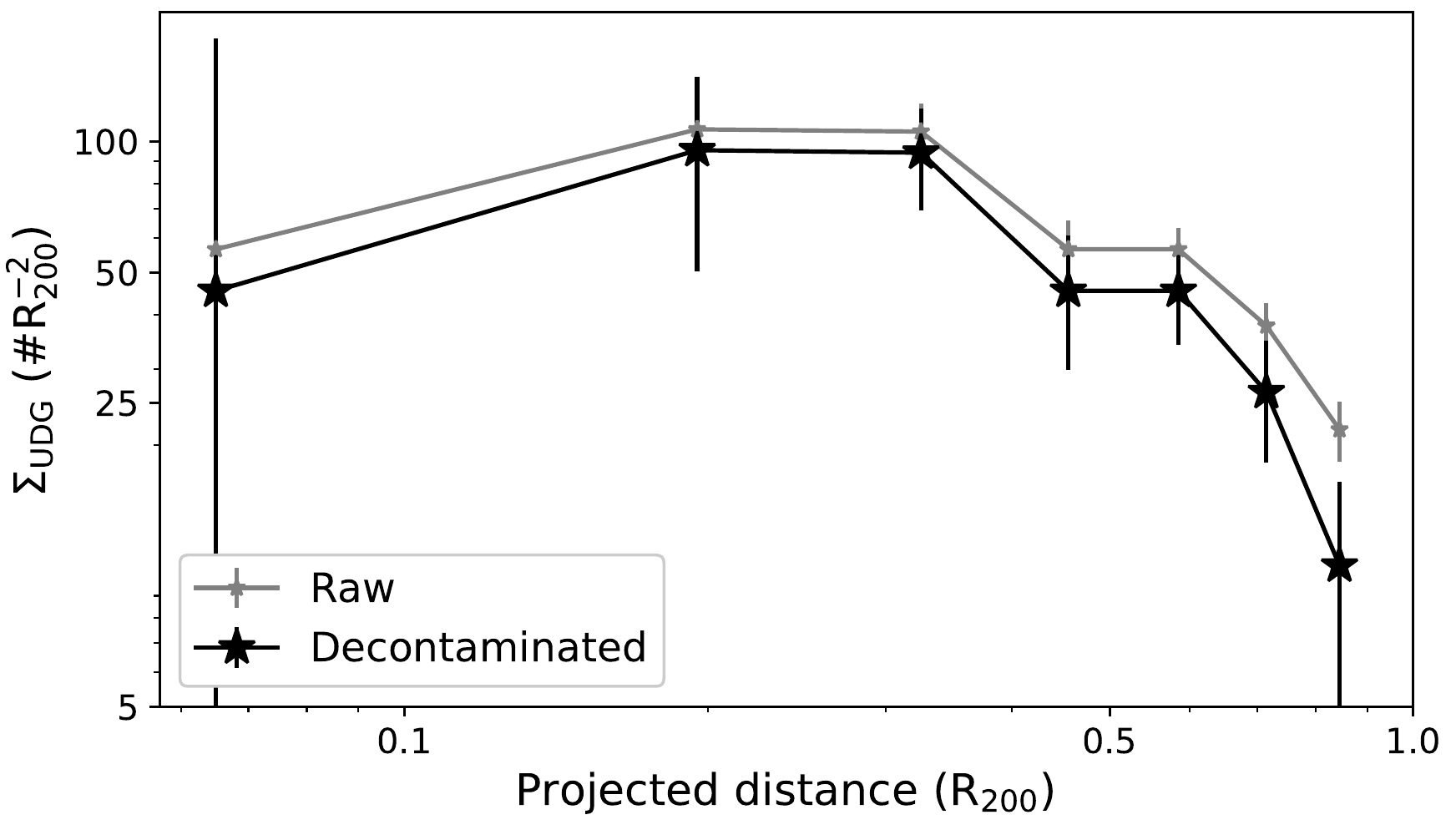}
\caption{The radial surface density profile of our cluster UDGs. Raw (grey, dashed) and decontaminated (black, solid) profiles. See the text for details.}
\label{fig:sdprof}
\end{figure}

We also study the spatial distribution of UDGs via their radial surface density profile as follows.
We estimate the surface density of UDGs in each cluster, as the number of UDGs in each bin of R$_{200}$ divided by the area of each bin, and decontaminating the profile from the expected background contamination. As explained in Paper I, the decontamination is performed by using observations of a blank field (observed in the same way as the clusters, and whose galaxies were analyzed following the same procedure as our sample, using \texttt{SExtractor} and \texttt{GALFIT}), counting the number of blank-field objects that would have been classified as UDGs in each cluster. The decontaminated number of UDGs is found by subtracting the expected contribution of interlopers from the original number of UDGs found in each cluster. We do not apply radial completeness corrections since the clusters are not strongly dominated by a bright cluster galaxy that could be hiding several UDGs (see also the discussion in \citealt{Aku}). The raw profile is shown in the top panel of Figure \ref{fig:sdprof} in grey, while the decontaminated profile is in black.

We checked and, similarly as in vdB+16, an Einasto profile \citep{einasto} provides a reasonable fit to the distribution of our data, with a profile that rises steeply inwards from (at least) 1 R$_{200}$ towards the inner parts, then becoming flat or developing a shallow core in the inner bins. These profiles are used in Paper I to show the differences between the expected and observed positions of the innermost UDGs, supporting a scenario were UDGs are destroyed in the centres of clusters.

\subsection{Projected clustercentric distance dependencies}
The projected distance is often used as a proxy of density within virialized clusters. Given the regular Gaussian-like velocity distribution of the galaxies in our clusters (Figure \ref{fig:redshifts}), as well as from visual inspection of X-ray contour maps, using the projected distance as a proxy of the density is reasonable, and it is interesting to see whether or not the properties of UDGs depend on it.
For instance, RT17b reported a decrease in stellar mass and effective radius of UDGs as they lie closer to the center of their host structures. Moreover, their galaxies with blue colors are also at larger projected distances than those with redder colors, a trend also confirmed in UDGs in Coma \citep{alabi}. Additionally, \cite{Aku3} found that the whole dwarf galaxy population in the FDS becomes slightly redder towards the center of the cluster, and that the early-type dwarf galaxies become redder in their $u'-X$ colors towards the center, whereas their $g-r$, $g-i$, and $r-i$ colors do not show significant trends.

On the other hand, by studying nearby clusters up to $z <$ 0.1, \cite{sanchez2} showed that within cluster environments the relatively red population of dwarf galaxies (like most of our UDGs) does not significantly change color as a function of projected clustercentric distance, as opposed to their bluer counterparts, which become significantly redder when approaching the cluster centres. 

To investigate the possible trends in our sample we first look at the colors, (non-circularized) effective radii, S\'ersic indices and axis ratios of our sample UDGs as a function of the projected distance. We do this for each cluster up to 1 R$_{200}$ to have full spatial coverage. While this does not reveal clear nor continuous trends the picture changes when we bin our data as follows.

\begin{figure*}
\centering
\includegraphics[scale=0.43]{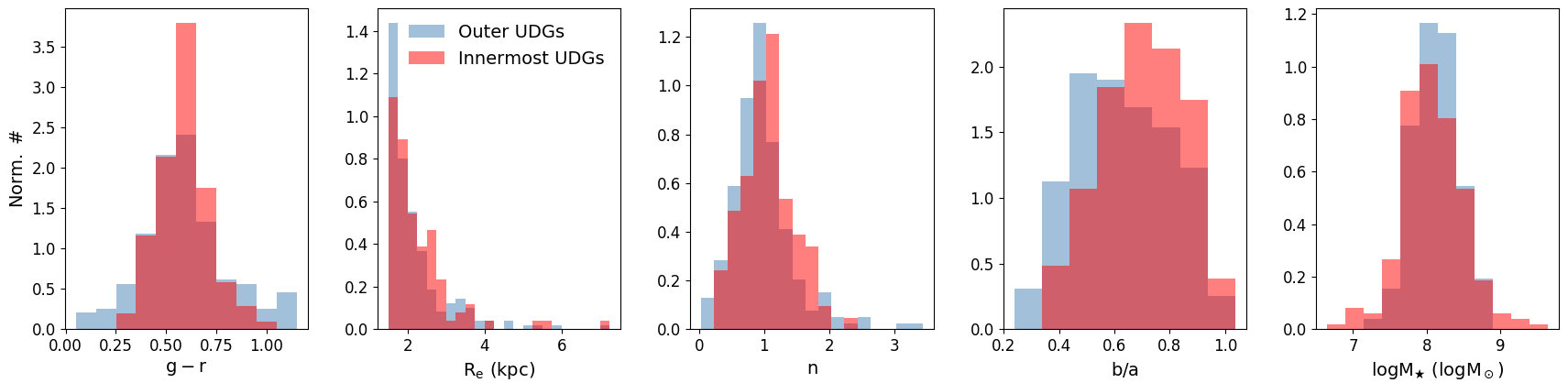}
\caption{Distribution of the color, effective radius, S\'ersic index, axis ratio and stellar mass for the innermost (inside 0.5 R$_{200}$) and outermost (outside 1 R$_{200}$) UDGs.}
\label{fig:histdist}
\end{figure*}

We split our sample in two extreme groups: those UDGs inside 0.5 R$_{200}$ and those outside 1 R$_{200}$. The idea is, as before, that while not in isolation, UDGs outside R$_{200}$ are relatively more isolated than the UDGs inside 0.5 R$_{200}$. We exclude the middle group in our analysis to increase the contrast between the other two groups (but see Sections \ref{sec:baReplane} and \ref{sec:colorReplane}).
Figure \ref{fig:histdist} shows the histograms of the structural parameters and the stellar mass for both groups. Qualitatively, the distributions of effective radius, and stellar mass look very similar, and the distributions of color, S\'ersic index and particularly of the axis ratios do look different. 

A K-S test comparing the two distributions on each histogram gives $p$-values of 0.31 for the effective radius and 0.09 for the stellar mass, for which we cannot reject the null hypothesis. For the S\'ersic index and the axis ratio, however, the $p$-values are 0.03 and 0.003, respectively, showing that inner and outer UDGs seem to have statistically significant different distributions of these two parameters: UDGs near the centres are, on average, more concentrated and rounder. Our discovery of a variation in S\'ersic index with clustercentric radius is similar to the result of \citealt{trujilloaguerri}, since they found that more-concentrated elliptical galaxies inhabit higher-density cluster regions than their less-concentrated counterparts.

The case of the color distributions requires more analysis because even when the $p$-value is 0.34, it is visible from Figure \ref{fig:histdist} that the distributions are not very similar; particularly in the outer parts of clusters there seem to be an excess of blue galaxies over the inner parts\footnote{A caveat to keep in mind for the rest of this discussion is that the larger spread in the colors of outer UDGs could be caused by a higher presence of interlopers than in the regions closer to the centres of clusters.}. To study this, we calculate the blue-to-red fractions of inner and outer UDGs. To separate between blue and red galaxies we use a threshold of $g-r < 0.5$ mag, motivated by the clear segregation between early- and late-type dwarfs in the FDS at that color \citep{Aku3}. We find a blue-to-red fractions of 0.32 for the innermost UDGs and 0.47 for the outer ones, indicating that the population of UDGs in lower-density environments has a higher contribution of blue galaxies than UDGs in higher-density regions. Moreover, to quantify 1$\sigma$ lower- (LL) and upper-limits (UL) to the contribution of blue galaxies to the innermost and outer UDGs we use the equations:
\begin{equation}
LL = \dfrac{Nb_{in,out} - Nb_{in,out}^{0.5}}{Nt_{in,out} - Nb_{in,out}^{0.5}},
\end{equation}
\begin{equation}
UL = \dfrac{Nb_{in,out} + Nb_{in,out}^{0.5}}{Nt_{in,out} + Nb_{in,out}^{0.5}}, 
\end{equation}
where $Nt$ denotes the total number of UDGs in the innermost and outer groups, and $Nb$ the number of blue galaxies inside each group.

Using this, we find limit values of 0.20--0.28 for the contribution of blue galaxies to the innermost UDGs, and 0.29--0.34 for its contribution to the outer UDGs. 
It is remarkable that we find different blue-to-red fractions and significant 1$\sigma$ differences for the upper and lower limits of the color distribution considering that i) the analysis has not been done in bins of mass, so galaxies of different masses are all mixed in our groups, ii) projection effects should be present when comparing the inner and outer UDGs, iii) UDGs are likely to be a mixed group or bag of galaxies, and iv) while expected to be low, we have some degree of background contamination. All these points add noise and scatter to the relation, so the fact that even with these sources of additional scatter we find 1$\sigma$-significance relations suggests that in practice the effects could be rather strong. Finally, it is important to clarify one point: based on the color distributions shown in Figure \ref{fig:histdist}, it could be claimed that the outer UDGs have also a higher fraction of red galaxies that
n the innermost UDGs. While this is true in principle, this trend is dominated by the contribution of the 10 galaxies showing the reddest colors ($g-r > 1$) of our sample, which, interestingly, are all part of the outer UDGs; from inspecting them we realize they are close to the edge of our mosaics, so their photometry could not be ideal, particularly in the $g$-band, and it is likely that the very red colors are not fully representative of these galaxies. Additionally, as already mentioned, the currently-used definitions of UDGs are likely to be missing blue counterparts, so one would expect the contribution of blue galaxies to increase, and since such younger UDGs are probably mostly found outside clusters this would show up preferentially in the outer regions. The final test would be for sure testing the distributions with more accurate colors and redshifts, but our data already seems to be in agreement with \cite{roman} and \cite{alabi}, finding a higher contribution of red UDGs with smaller projected clustercentic distance.

Overall, our observations of the S\'ersic index of galaxies increasing towards the inner regions while becoming also rounder are in good agreement with expectations from models of dwarf galaxies that have undergone harassment and tidal interaction processes (\citealt{moore}, \citealt{aguerriygg}, and see also \citealt{lisker}).
The higher contribution of blue galaxies to the outer group compared with the contribution to the innermost group is as expected from an extension of the morphology-density relation (e.g. \citealt{dressler}), and it is in excellent agreement with the trends found for dwarf galaxies in Fornax \citep{Aku3}, where the blue-to-red fraction of galaxies increases for larger clustercentric distances, as well as for a large number of low-redshift dwarfs in \cite{sanchez2}.

\begin{figure*}
\centering
\includegraphics[scale=0.45]{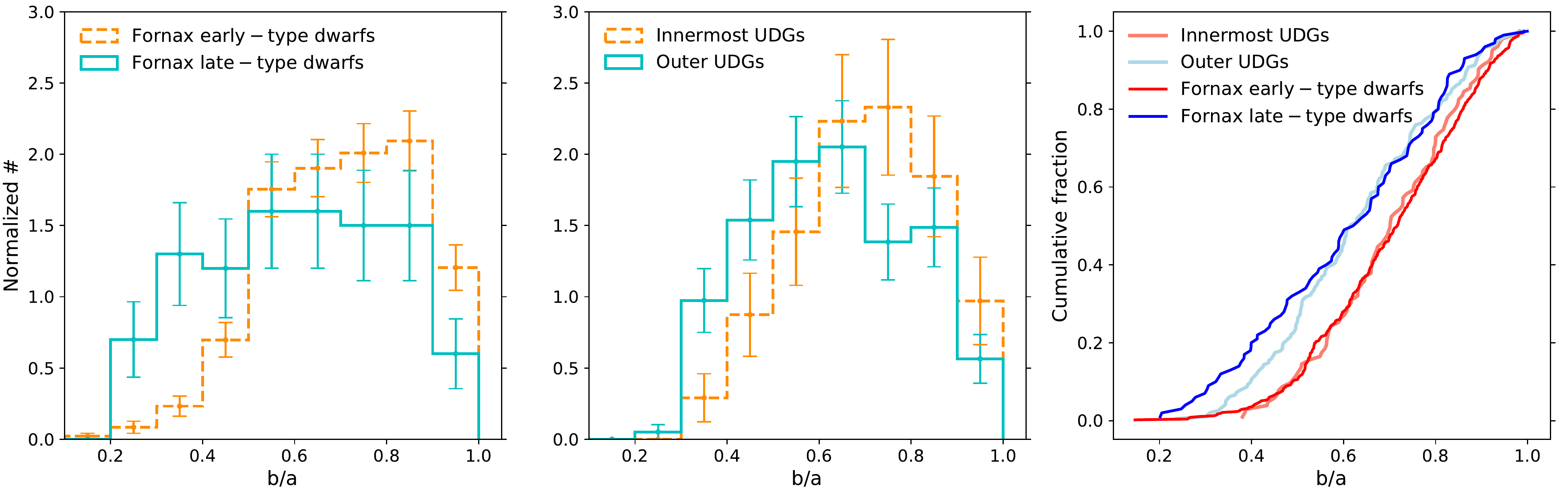}
\caption{\textit{Left}: Axis ratio distribution of late-type (cyan) and early-type (orange) dwarfs in the Fornax Cluster \citep{Aku3}. \textit{Middle}: Axis ratio distribution of outer (cyan) and innermost (orange) UDGs; the errorbars in both panels are the normalized Poissonian uncertainties. \textit{Right:} Cumulative fraction of the four groups of galaxies. The distribution of late-type dwarfs is similar to the distribution of outer UDGs, as the distributions of early-type dwarfs and innermost UDGs are. See the text for details.}
\label{fig:ARs}
\end{figure*}

Furthermore, the result that the axis ratio distributions depends on clustercentric distance takes more importance when compared with the distribution of dwarf galaxies from the FDS \citep{Aku3}. In that work, the authors studied deep photometric observations of dwarf galaxies, classifying them as either late-type dwarfs or dEs (see also Figure \ref{fig:scaling}). In Figure \ref{fig:ARs} we show the axis ratio distribution for late-type (star forming) and early-type (quiescent) dwarfs in Fornax, as well as our distribution of inner and outer UDGs, and the cumulative fractions of the four groups of galaxies. A striking similarity appears when comparing both works: the axis ratio distribution of our innermost UDGs follows remarkably the distribution of the early-type dwarfs in Fornax, and the distribution of the outer UDGs (specially for $b/a > 0.5$) the one of the Fornax late-type dwarfs. This suggests an evolutionary scenario were UDGs are being transformed when approaching to the centre of clusters, just as late-type dwarfs are transformed by the environment into dEs/dSphs. 

\subsection{Host cluster mass dependences}
The total mass of clusters acts like a global environmental proxy, and it is interesting to study if the structural parameters of UDGs change systematically in clusters of different masses.
For instance, galaxies inhabiting clusters with low $\sigma$, that are less massive, are expected to have undergone stronger galaxy-galaxy interactions than galaxies in clusters with high $\sigma$, since the low-velocities increase the cross section for mergers (e.g. \citealt{lefevre}).
On the other hand, more-massive clusters have stronger potentials and ram-pressure stripping (which goes as $\rho \sigma^2$, with $\rho$ the gas density) is extremely strong in them (e.g. \citealt{gunn}).

\begin{figure}
\centering
\includegraphics[scale=0.42]{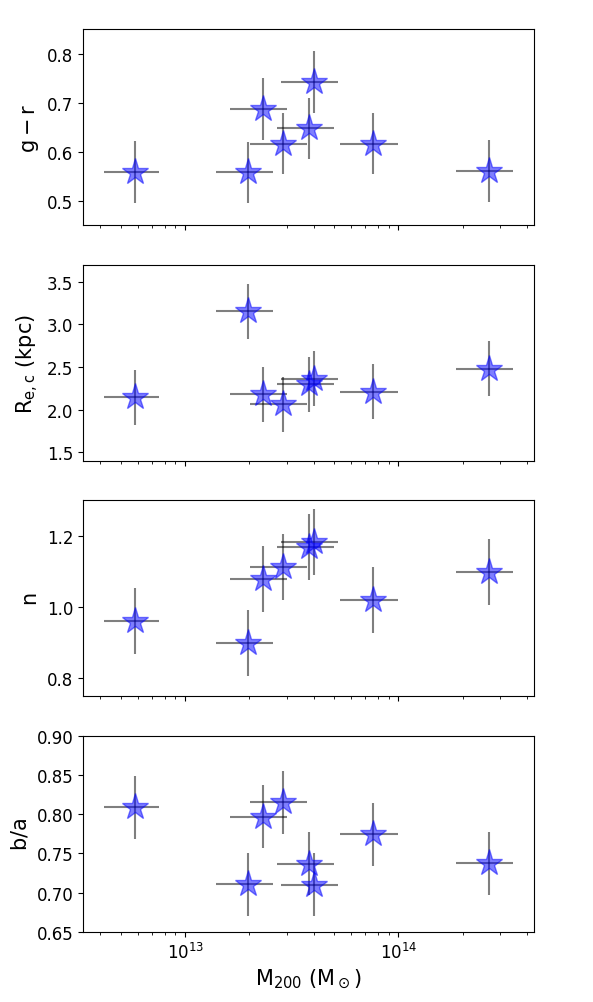}
\caption{Mean structural parameters of UDGs in each cluster as a function of the M$_{200}$ of the host cluster. The y-uncertainties are the standard deviation of each parameter on each cluster.}
\label{fig:massdep}
\end{figure}

To investigate if any dependence on host cluster mass is present, we look at the mean values of the structural parameters as a function of M$_{200}$. Figure \ref{fig:massdep} shows these mean values for our UDG candidates in each cluster.

Examining our data we do not see any evident radial trend for the studied parameters. The distributions are particularly flat for the color and effective radius, while for the S\'ersic index and axis ratio it seems to be some inkling of a trend but it is not clear with our data. If real, these trends would suggest that UDGs inhabiting more massive clusters have on average more concentrated surface brightness profiles and lower axis ratios.

These trends could have a physical origin: galaxy harassment can increase the S\'ersic index in disk galaxies, such as UDGs (but it should also increase the axis ratio, which is not clearly observed), and stripping would make galaxies in high-mass systems become more disrupted and have lower axis ratios. 

Giving our relatively small sample we are not in position of giving a conclusive answer on whether or not the cluster mass M$_{200}$ systematically affects the population of UDGs, but it is for sure something that should be studied with more data and will be one of the goals once \texttt{KIWICS} is complete.

\section{Conclusions}
\label{sec:conclusions}

In this work we studied the population of UDGs in a set of eight nearby galaxy clusters from the \texttt{KIWICS} sample. To summarise, the main findings of our study are:
\begin{itemize}
\item We find 442 new UDG candidates, 247 of them lying at projected distances < 1 R$_{200}$ of their associated cluster. They have mainly red colors of passively evolving stellar populations, although they appear in a range in colors. Large UDGs are rare, and the distribution is dominated by UDGs with $R_e \leq$ 2.5 kpc. They have basically exponential light profiles and stellar masses $\sim$10$^8$ M$\odot$ (e.g. Figure \ref{fig:histogram}).

\item Overall, they follow the behavior of dwarf galaxies in different scaling relations, standing out only for their larger size. While their spread in colors is relatively high, the bulk of them fit well the RS. The color-magnitude diagram also highlights the fact that UDGs should be a mix of early- and late-type dwarfs: the separate sequences of early-and late-type dwarfs becomes a cloud when UDGs appear. This is also supported for the different axis ratio distributions that inner and outer UDGs follow (Figure \ref{fig:scaling}).

\item We find no systematic evidence of the size of UDGs depending on their color, but inside clusters small UDGs are rounder than large UDGs (Figure \ref{fig:ARCOLvsRe}). If colors of UDGs inside clusters are still indicatives of their SFHs our observations would imply that the different sizes of UDGs do not depend on the SFHs, as expected in the model by \cite{dicintio}. Additionally we do not observe large UDGs being systematically redder than small UDGs, as proposed by \cite{carleton}. A caveat that should be kept in mind is the uncertainties in our color determinations.


\item It seems that radial galaxy alignment is not a common feature of cluster UDGs. Only in one of our eight clusters the UDGs have a distribution of angles relative to the center compatible with not being flat.

\item There are indications that the contribution of blue UDGs in the outer regions of clusters is higher than in the innermost regions (see also RT17b and \citealt{alabi}). Additionally, UDGs in the innermost regions have on average slightly higher S\'ersic indices and larger axis ratios than relatively isolated UDGs (Figure \ref{fig:histdist}), in agreement with a scenario where harassment plays an important role shaping the galaxies. Moreover, the axis ratio distribution of these relatively isolated (outer) UDGs resembles the distribution of late-type dwarfs, while innermost UDGs mirror early-type dwarfs (Figure \ref{fig:ARs} and see also \citealt{Aku3}). This suggests that UDGs are being transformed in clusters in the same way as other dwarf galaxies.
\end{itemize}

Overall, our findings favor a picture in which UDGs are dwarf-like galaxies accreted from the field or smaller groups to clusters. During this process they follow a relatively passive evolution where the cluster environment quenches their star formation and they experience harassment and tidal disrupting forces. As a result, UDGs would become redder, rounder and more concentrated towards the centres of clusters, resembling the transformation of late-type to early-type dwarfs.

\section*{Acknowledgements}
This work is based on observations made with the Isaac Newton Telescope operated on the island of La Palma by the Isaac Newton Group of Telescopes in the Spanish Observatorio del Roque de los Muchachos of the Instituto de Astrof\'isica de Canarias. 

A careful revision and valuable comments by an anonymous referee are highly appreciated. We thank Javier Rom\'an for the data of RT17b and for enlightening comments and discussions about our analysis and results. We thank also Remco van der Burg for sharing the data of vdB+16 with us as well as for many clarifications on it. 

PEMP thanks the Netherlands Research School for Astronomy (NOVA) for funding via the NOVA MSc Fellowship. PEMP, RFP and AV acknowledge financial support from the European Union's Horizon 2020 research and innovation programme under Marie Sk\l odowska-Curie grant agreement No. 721463 to the SUNDIAL ITN network. JALA acknowledges support from the Spanish Ministerio de Econom\'ia y Competitividad (MINECO) by the grants AYA2013-43188-P and AYA2017-83204-P. AV would like to thank the Vilho, Yrj\"o, and Kalle V\"ais\"al\"a Foundation of the Finnish Academy of Science and Letters for the funding during the writing of this paper.

We have made an extensive use of SIMBAD and ADS services, as well as of the Python packages NumPy \citep{numpy}, Matplotlib \citep{hunter} and Astropy \citep{astropy}, for which we are thankful.





\begin{thebibliography}{199}
  \bibitem[Abolfathi et al.(2018)]{sdss} Abolfathi et al. 2018; ApJS, 235, 42A 
  \bibitem[Aguerri \& Gonz\'alez-Garc\'ia(2009)]{aguerriygg} Aguerri, J. A. L., \& Gonz\'alez-Garc\'ia, A. C. 2009, A\&A, 494, 891
  \bibitem[Agulli et al.(2014)]{agulli} Agulli, I., et al. 2014, MNRAS, 444, 34
  \bibitem[Alabi et al.(2018)]{alabi} Alabi, A., et al. 2018, MNRAS, 479, 3308
  \bibitem[Amorisco(2018)]{amoriscohaloes} Amorisco, N. C. 2018, MNRAS, 475, L116
  \bibitem[Amorisco \& Loeb(2016)]{amorisco} Amorisco, N. C., \& Loeb, A. 2016, MNRAS, 459, 51
  \bibitem[Amorisco et al.(2018)]{amorisco2018} Amorisco, N. C., Monachesi, A., \& White, S. D., 2018, MNRAS, 475, 4235 
  \bibitem[Astropy Collaboration et al.(2013)]{astropy} Astropy Collaboration et al. 2013, A\&A, 558, A3
  \bibitem[Baushev(2018)]{baushev} Baushev, A. N., 2018, NewA, 60, 69B   
        \bibitem[Beasley et al.(2016)]{beasley} Beasley, M. A., et al. 2016, ApJ, 819, L20
      \bibitem[Beasley \& Trujillo(2016)]{beasleyytrujillo} Beasley, Michael A.; Trujillo, Ignacio; ApJ, 830, 26
        \bibitem[Bellazzini et al.(2017)]{bellazzini} Bellazzini et al. 2017, MNRAS, 467, 3751.
    \bibitem[Bennet et al.(2018)]{bennet} Bennet, P., et al., 2018, submitted to ApJL, \url{arXiv:1809.01145} 
        \bibitem[Bertini \& Arnouts(1996)]{sextractor} Bertin, E., \& Arnouts, S. 1996, A\&AS, 117, 393
    \bibitem[Bilir, Karaali \& Tun{\c c}el(2005)]{bilir} Bilir S., Karaali S., Tun{\c c}el(2005) S., 2005, Astron. Nachr., 326, 321
    \bibitem[Bothun et al.(1991)]{bothun} Bothun G. D., Impey C. D., Malin D. F., 1991, ApJ, 376, 40
    \bibitem[Carleton et al.(2018)]{carleton} Carleton, T., et al. 2018; submitted to MNRAS, \url{arXiv: 1805.06896 }
  \bibitem[Chabrier(2003)]{chabrier} Chabrier, G. 2003, PASP, 115, 763
    \bibitem[Chilingarian \& Zolotukhin(2012)]{chilingarian} Chilingarian, I., Zolotukhin, I., 2012, MNRAS, 419, 1727m
    \bibitem[Cohen et al.(2018)]{cohen} Cohen, Y., et al. 2018, submitted to ApJ, \url{arXiv:1807.06016}
    \bibitem[Conselice(2018)]{conselice} Conselice C. J., 2018, RNAAS, 2, 43
      \bibitem[Cutri et al.(2003)]{2mass} Cutri, R. M., Skrutskie, M. F., van Dyk, S., et al. 2003, VizieR Online Data Catalog, 2246
      \bibitem[Dalcanton et al.(1997a)]{dalcanton} Dalcanton, J. et al. 1997a, AJ, 114, 635
      \bibitem[Dalcanton et al.(1997b)]{dalcantondisks} Dalcanton, J. et al. 1997b, ApJ, 482, 659
      \bibitem[Dalton et al.(2016)]{dalton} Dalton, G., Trager, S., et al. 2016, Porc. SPIE, 9908, 1
   \bibitem[Di Cintio et al.(2017)]{dicintio} Di Cintio, A., Brook, C. B., Dutton, A. A., et al. 2017, MNRAS, 466, L1
   \bibitem[Dressler(1980)]{dressler} Dressler, A. 1980, ApJ, 236, 351
   \bibitem[Einasto(1965)]{einasto} Einasto J., 1965, Trudy Astrofizicheskogo Instituta Alma-Ata, 5, 87
     \bibitem[Ferr\'e-Mateu et al.(2018)]{ferre} Ferr\'e-Mateu, A., et al. 2018, MNRAS, 479, 4891 
       \bibitem[Gunn \& Gott(1972)]{gunn} Gunn, J. E., \& Gott, J. R., III. 1972, ApJ, 176, 1
       \bibitem[Greco et al.(2018)]{greco} Greco, J., et al. 2018, ApJ, 866, 112
       \bibitem[Holwerda(2005)]{sexford} Holwerda, B. W., 2005, arXiv:astro-ph/0512139
      \bibitem[Hunter(2007)]{hunter} Hunter J. D., 2007, Computing In Science \& Engineering, 9, 90
       \bibitem[Impey \& Bothun(1997)]{reviewlsb} Impey, C. \& Bothun, G., 1997, ARA\&A, 35, 267
       \bibitem[Impey et al.(1988)]{impey} Impey C., Bothun G., Malin D., 1988, ApJ, 330, 634
          \bibitem[Jones et al.(2018)]{jones} Jones, M. G., et al. 2018, A\&A, 614, 21
       \bibitem[Koda et al.(2015)]{koda} Koda, J., Yagi, M., Yamanoi, H., \& Komiyama, Y., 2015, ApJL, 807, L2
  \bibitem[Lim et al.(2018)]{lim} Lim, S., et al. (2018), accepted in ApJ
  \bibitem[Lisker et al.(2006)]{lisker} Lisker, T., Glatt, K., Westera, P., \& Grebel, E. 2006, AJ, 132, 24
    \bibitem[Le F\'{e}vre et al.(2000)]{lefevre} Le F\'{e}vre, O., et al. 2000, MNRAS, 311, 565
    \bibitem[Leisman et al.(2017)]{leisman} Leisman L., et al. 2017, ApJ, 842, 13
    \bibitem[Mancera Pi\~na et al.(2018)]{paperI} Mancera Pi\~na, P. E., Peletier, R. F., et al. 2018, MNRAS, 481, 4381
       \bibitem[Martin et al.(2018)]{martin} Martin, N. F. et al. 2018, ApJL, 859, 5
     \bibitem[McFarland et al.(2013)]{astrowise} McFarland, J. P., Verdoes-Kleijn,  G., Sikkema, G., et al. 2013, Experimental Astronomy, 35, 45 
    \bibitem[Mihos et al.(2015)]{mihos} Mihos et al. 2015, ApJ, 809, 21
      \bibitem[Moore et al.(1996)]{moore} Moore, B., et al. 1996, Nature, 379, 613
      \bibitem[Munari et al.(2013)]{munari} Munari E., Biviano A., Borgani S., Murante G., Fabjan D., 2013, MNRAS, 430, 2638
        \bibitem[Oliphant(2006)]{numpy} Oliphant, T. 2006, A Guide to NumPy, 1 (Trelgol Publishing USA
     \bibitem[Papastergis et al.(2017)]{papastergis} Papastergis, E., Adams, E. A. K., \& Romanowsky, A. J. 2017, A\&A, 601, L10
       \bibitem[Peng(2010)]{galfit} Peng, C. Y., et al. 2010, AJ, 139, 2
      \bibitem[Piffaretti et al.(2011)]{piffaretti} Piffaretti R., Arnaud M., Pratt G. W., Pointecouteau E., Melin J. B., 2011, A\&A, 534, A10
      \bibitem[Posti et al.(2018)]{posti} Posti, L., et al. 2018, MNRAS, 475, 232
      \bibitem[Roediger \& Courteau(2015)]{m2l} Roediger, J. C., \& Courteau, S. 2015, MNRAS, 452, 3209

        \bibitem[Rom\'an \& Trujillo(2017a)]{roman} Rom\'an, J., \& Trujillo, I. 2017a, MNRAS, 468, 703
    \bibitem[Rom\'an \& Trujillo(2017b)]{roman2} Rom\'an, J., \& Trujillo, I. 2017b, MNRAS, 468, 4039 
  
    \bibitem[Ruiz-Lara et al.(2018)]{ruizlara} Ruiz-Lara, T., et al. 2018, MNRAS, 478, 2034
        \bibitem[S\'anchez-Janssen(2009)]{sanchez} S\'anchez-Janssen, R., 2009, PhD thesis, Universidad de la Laguna
        \bibitem[S\'anchez-Janssen et al.(2008)]{sanchez2} S\'anchez-Janssen, R., Aguerri, J. A. L., \& Mu\~noz-Tu\~n\'on, C. 2008, ApJ, 679, L
        \bibitem[Sandage \& Binggeli(1984)]{sandage} Sandage \& Binggeli, 1984, AJ, 89, 919
                \bibitem[Schlafly \& Finkbeiner(2011)]{schlafly} Schlafly, E.F., \& Finkbeiner, D.P., 2011, ApJ, 737, 103
                \bibitem[S\'ersic(1963)]{sersic} S\'ersic, J. L. 1963, Bolet\'in de la Asociaci\'on Argentina de Astronom\'ia La Plata Argentina, 6, 4
          \bibitem[Taylor(2005)]{topcat} Taylor, M. B. 2005, Astronomical Data Analysis Software and Systems XIV, 347
      \bibitem[Tolman(1930)]{tolman1} Tolman, R. C., 1930, Proceedings of the National Academy of Science, 16, 511
    \bibitem[Tolman(1934)]{tolman2} Tolman, R. C., 1934, Relativity, Thermodynamics, and Cosmology.
    \bibitem[Toloba et al.(2018)]{toloba} Toloba et al. 2018, ApJ, 856, 31
      \bibitem[Trujillo et al.(2002)]{trujilloaguerri} Trujillo, I., et al. 2002, ApJ, 573, L9
      \bibitem[Trujillo et al.(2017)]{trujillo} Trujillo, I., Rom\'an, J., Filho, M., \& S\'anchez Almeida, J. 2017, ApJ, 836, 191
  \bibitem[van der Burg et al.(2016)]{vanderburg} van der Burg, R. F. J., Muzzin, A., \& Hoekstra, H. 2016, A\&A, 590, A20
  \bibitem[van Dokkum et al.(2015)]{vandokkum} van Dokkum, P. G., et al. 2015, ApJL, 798,  L45
    \bibitem[van Dokkum et al.(2016)]{vandokkum2} van Dokkum P., et al. 2016, ApJ, 828, L6   
  \bibitem[Venhola et al.(2017)]{Aku} Venhola, A., et al. 2017, A\&A, 608, 142
    \bibitem[Venhola et al.(2018a)]{Aku2} Venhola, A., et al. 2018a, to appear in A\&A
    \bibitem[Venhola et al.(2018b, submitted)]{Aku3} Venhola, A., et al. 2018b, submitted to A\&A
   \bibitem[Wang et al.(2015)]{nihao} Wang, L., et al. 2015, MNRAS, 454, 83
      \bibitem[Yagi et al.(2016)]{yagi} Yagi, M., Koda, J., Komiyama, Y., \& Yamanoi, H. 2016, ApJS, 225, 11
\bibitem[Yozin \& Bekki(2015a)]{yozin} Yozin, C., \& Bekki, K. 2015a, MNRAS, 452, 9

\bibitem[Yozin \& Bekki(2015b)]{yozin2} Yozin, C., \& Bekki, K. 2015b, MNRAS, 453, 14
\end{thebibliography}


\appendix
\onecolumn
\section{\texttt{SExtractor} configuration}
\label{sec:A2}
In this appendix we provide the non-default \texttt{SExtractor} parameters used for each of our clusters. The configuration file with these parameters was given to \texttt{SExtractor} to run in the dual mode, using the $r-$band image to detect the sources.
\begin{itemize}
\item RXCJ1714: \texttt{DETECT\_MINAREA = 20, DETECT\_THRESHOLD = 1.5, ANALYSIS\_THRESHOLD = 1.5}.
\item RXCJ1223: \texttt{DETECT\_MINAREA = 20, DETECT\_THRESHOLD = 1.3, ANALYSIS\_THRESHOLD = 1.3}.
\item MKW4S: \texttt{DETECT\_MINAREA = 20, DETECT\_THRESHOLD = 1.4, ANALYSIS\_THRESHOLD = 1.4}.
\item RXCJ1204: \texttt{DETECT\_MINAREA = 20, DETECT\_THRESHOLD = 1.3, ANALYSIS\_THRESHOLD = 1.3}.
\item A1177: \texttt{DETECT\_MINAREA = 20, DETECT\_THRESHOLD = 1.5, ANALYSIS\_THRESHOLD = 1.5}.
\item A779: \texttt{DETECT\_MINAREA = 20, DETECT\_THRESHOLD = 1.5, ANALYSIS\_THRESHOLD = 1.5}.
\item A1314: \texttt{DETECT\_MINAREA = 20, DETECT\_THRESHOLD = 1.3, ANALYSIS\_THRESHOLD = 1.3}.
\item A2634: \texttt{DETECT\_MINAREA = 20, DETECT\_THRESHOLD = 1.4, ANALYSIS\_THRESHOLD = 1.4}.
\end{itemize}

\section{Structural parameters of UDGs in each galaxy cluster}
\label{sec:A1}
\begin{table*}
\caption{Mean, median, minimum and maximum value, and the dispersion of each structural parameter in the inner 1 R$_{200}$ of each cluster. The order of the clusters here is different that in previous tables: here they are ordered from low to high masses. For reference, last column indicates the M$_{200}$ of each cluster.} 
\label{tab:params}
\begin{center}
\begin{tabular}{llcccccc}
\hline
\noalign{\smallskip}
    \texttt{$g-r$} \\ \hline     \noalign{\smallskip}
   & Cluster & mean & median & min & max & $\sigma$ & M$_{200}$ ($\times$10$^{13}$ M$_\odot$)\\ \hline
    \noalign{\smallskip} 
&RXCJ1714  &  0.58 &  0.56 &  0.50 &  0.66 &  0.05 & 0.58\\ \noalign{\smallskip}
 &RXCJ1223  &  0.55 &  0.55 &  0.45 &  0.65 &  0.05 & 1.98 \\ \noalign{\smallskip}
  &  MKW4S  &  0.58 &  0.62 &  0.30 &  0.98 &  0.17 & 2.31 \\ \noalign{\smallskip}
 &RXCJ1204  &  0.59 &  0.55 &  0.36 &  0.92 &  0.16 & 2.88 \\ \noalign{\smallskip}
  &  A1177  &  0.64 &  0.64 &  0.46 &  0.84 &  0.12 & 3.82\\ \noalign{\smallskip}
   & A779   &  0.74 &  0.72 &  0.52 &  0.94 &  0.13 &4.02 \\ \noalign{\smallskip}
   & A1314  &  0.63 &  0.65 &  0.37 &  0.91 &  0.12 & 7.62\\ \noalign{\smallskip}
&	A2634  &  0.56 &  0.57 &  0.22 &  0.89 &  0.13 & 26.60  \\ 
	\hline
	\noalign{\smallskip}
    \texttt{$R_e$ (kpc)} \\ \hline     \noalign{\smallskip}
 &RXCJ1714  &  2.09 &  1.77 &  1.64 &  3.32 &  0.55 &0.58\\ \noalign{\smallskip}
 &RXCJ1223  &  2.62 &  2.34 &  1.52 &  7.04 &  1.33&1.98\\ \noalign{\smallskip}
 &   MKW4S  &  1.85 &  1.82 &  1.53 &  2.47 &  0.32 & 2.31 \\ 
 &RXCJ1204  &  1.97 &  1.91 &  1.57 &  2.98 &  0.33 & 2.88\\ \noalign{\smallskip}
 &   A1177  &  2.08 &  1.93 &  1.53 &  3.63 &  0.52 & 3.82\\ \noalign{\smallskip}
  &  A779   &  2.26 &  2.03 &  1.51 &  3.95 &  0.64 &4.02\\ \noalign{\smallskip}
   & A1314  &  1.97 &  1.80 &  1.51 &  3.35 &  0.45 & 7.62\\ \noalign{\smallskip}
&	A2634  &  2.13 &  1.96 &  1.50 &  5.70 &  0.68& 26.60\\ 
	\noalign{\smallskip}
	\hline
    \texttt{n} \\ \hline     \noalign{\smallskip}
 &RXCJ1714  &  0.93 &  0.79 &  0.62 & 1.60 &  0.31 &0.58\\ \noalign{\smallskip}
 &RXCJ1223  &  0.93 &  0.95 &  0.48 & 1.35 &  0.20&1.98\\ \noalign{\smallskip}
  &  MKW4S  &  0.98 &  0.95 &  0.04 & 1.65 &  0.40 & 2.31 \\ \noalign{\smallskip}
 &RXCJ1204  &  1.03 &  1.04 &  0.55 & 1.75 &  0.31 & 2.88\\ \noalign{\smallskip}
  &  A1177  &  1.14 &  1.14 &  0.58 & 1.89 &  0.31 & 3.82\\ \noalign{\smallskip}
   & A779   &  1.14 &  1.06 &  0.24 & 2.98 &  0.56 &4.02\\ \noalign{\smallskip}
   & A1314  &  1.03 &  1.05 &  0.37 & 1.83 &  0.39 & 7.62\\ \noalign{\smallskip}
&	A2634  &  1.03 &  0.98 &  0.17 & 2.33 &  0.46& 26.60 \\ \hline
    \texttt{b/a} \\
     \hline     \noalign{\smallskip}
 &RXCJ1714  &  0.79 &  0.82 &  0.62 &  0.89 &  0.10 &0.58\\ \noalign{\smallskip}
 &RXCJ1223  &  0.71 &  0.72 &  0.38 &  0.98 &  0.16 &1.98\\ \noalign{\smallskip}
 &   MKW4S  &  0.68 &  0.65 &  0.38 &  0.92 &  0.16 & 2.31 \\ \noalign{\smallskip}
 &RXCJ1204  &  0.74 &  0.77 &  0.36 &  0.95 &  0.15 & 2.88\\ \noalign{\smallskip}
  &  A1177  &  0.73 &  0.70 &  0.60 &  0.93 &  0.11 & 3.82\\ \noalign{\smallskip}
   & A779   &  0.71 &  0.69 &  0.38 &  0.96 &  0.17 &4.02\\ \noalign{\smallskip}
   & A1314  &  0.71 &  0.73 &  0.38 &  0.97 &  0.15 & 7.62\\ \noalign{\smallskip}
&	A2634  &  0.67 &  0.68 &  0.33 &  0.97 &  0.16 & 26.60\\ 
      \hline
	\end{tabular} 
	\end{center}
	\end{table*}

%






\bsp	
\label{lastpage}
\end{document}